\documentclass[a4paper,english,onecolumn, draftcls, 12pt]{IEEEtran}
\usepackage[T1]{fontenc}
\usepackage[latin9]{inputenc}
\usepackage{verbatim}
\usepackage{amsmath}
\usepackage{graphicx}
\usepackage{setspace}
\usepackage{amssymb}
\makeatletter

\providecommand{\tabularnewline}{\\}

\makeatother

\usepackage{babel}

\begin{document}

\title{Partner Selection\\
in Indoor-to-Outdoor Cooperative Networks:\\
an Experimental Study%
\thanks{This paper was presented in part at the IEEE International Conference
on Communications (ICC 2010), Cape Town, South Africa, in May 2010.
This work has been submitted to the IEEE for possible publication.
Copyright may be transferred without notice, after which this version
may no longer be accessible.%
}}

\author{P. Castiglione\authorrefmark{1}, S. Savazzi\authorrefmark{2}, M.
Nicoli\authorrefmark{2}, T. Zemen\authorrefmark{1}\vspace{-3mm}\authorblockA{\\\authorrefmark{1}Forschungszentrum Telekommunikation Wien, Austria} \vspace{-3mm}\authorblockA{\\\authorrefmark{2}Dip. di Elettronica e Informazione, Politecnico di Milano, Italy}\vspace{-5mm}\vspace{-9.5mm}
}
\maketitle
\begin{abstract}
\thispagestyle{empty}In this paper, we develop a partner selection
protocol for enhancing the network lifetime in cooperative wireless
networks. The case-study is the cooperative relayed transmission from
fixed indoor nodes to a common outdoor access point. A stochastic
bivariate model for the spatial distribution of the fading parameters
that govern the link performance, namely the Rician K-factor and the
path-loss, is proposed and validated by means of real channel measurements.
The partner selection protocol is based on the real-time estimation
of a function of these fading parameters, i.e., the coding gain. To
reduce the complexity of the link quality assessment, a Bayesian approach
is proposed that uses the site-specific bivariate model as a-priori
information for the coding gain estimation. This link quality estimator
allows network lifetime gains almost as if all K-factor values were
known. Furthermore, it suits IEEE 802.15.4 compliant networks as it
efficiently exploits the information acquired from the receiver signal
strength indicator. Extensive numerical results highlight the trade-off
between complexity, robustness to model mismatches and network lifetime
performance. We show for instance that infrequent updates of the site-specific
model through K-factor estimation over a subset of links are sufficient
to at least double the network lifetime with respect to existing algorithms
based on path loss information only.\end{abstract}
\newpage{}

\section{Introduction}

Wireless sensor networks (WSN) are the enabling technology for home
and building automation \cite{Gom10,Industrial}. The main obstacle
in the development of WSNs is the cost of battery replacement, which
becomes even more pronounced for indoor-to-outdoor (I2O) communication
with its larger transmit powers requirements. In this paper we aim
at designing a medium access control (MAC) protocol for I2O WSNs that
minimizes the maximum transmit energy so as to \textit{\emph{prolong
 the network lifetime}} \cite{Chen05a}.

Transmit energy can be reduced by implementing advanced cooperative
relaying strategies \cite{Cui04a}, that efficiently exploit the inherent
spatial diversity of a distributed radio channel. Selecting the partner
for each node \cite{Hunter04} is the most crucial task in the coordination
phase of any cooperative technique \cite{Liu06}. \textit{\emph{Partner
selection}} is either based on instantaneous or average channel quality
indicators of the links \cite{Ble06,Mahinthan08}. Additional knowledge
on macroscopic features such as the network topology \cite{Zor03}
and a parametric characterization of the fading channel, e.g., a path-loss
model \cite{Lin06b}, can be also exploited.

\begin{figure}[t]
\centering{\includegraphics[width=12cm]{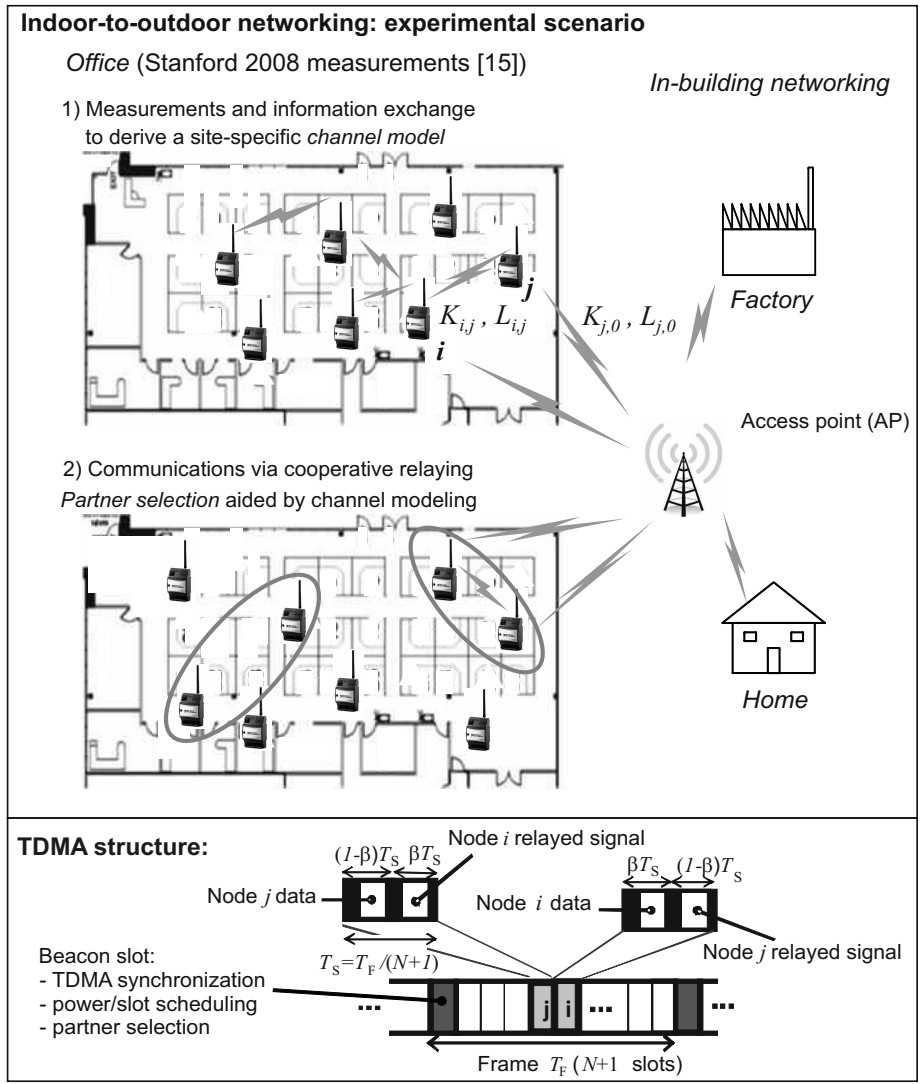}}

\caption{Top: general indoor networking scenario and the specific I2O office
radio measurement plan in \cite{Czink08}; indoor nodes are allowed
to support the estimation of a site-specific stochastic channel model
(1), and to engage in cooperative transmission to the access point
(2). Bottom: TDMA framing structure inspired to the IEEE 802.15.4
beacon mode.}

\label{Flo: intro}%
\end{figure}

In this paper we focus on the transmission from several indoor \textit{static}
nodes in a single-floor office or factory to a common outdoor access
point (AP), as outlined in Fig. \ref{Flo: intro} (top). Nodes are
permitted to engage in cooperative transmissions by amplifying and
forwarding (AF) \cite{Laneman04} the signals received from the partner
nodes. We are interested to develop a partner selection scheme that
maximizes the network lifetime under reliability and rate constraints
that are identical for all nodes. We propose an original approach
where the partner selection is aided by the knowledge of a site-specific
multi-link stochastic channel model. The network employs a time division
multiple access (TDMA) scheme inspired by the IEEE 802.15.4 access
protocol in {}``beacon mode'' \cite{Cal02}, as depicted in Fig.
\ref{Flo: intro} (bottom).

In \cite{Ble06} selection cooperation based on instantaneous channel
state information (CSI) has been introduced. For large networks such
an approach requires the exchange of large amounts of CSI within the
coherence time of the fading channel. To reduce the transmission overhead,
here we are interested to employ long-term channel properties for
the partner selection stage. This approach has been used by the pairing
protocol in \cite{Mahinthan08}, which is based on the path loss values.
This is a very practical choice, as e.g. in the IEEE 802.15.4 standard
a received signal strength indicator (RSSI) is available \cite{Srinivasan06a}.
However, an indication of the randomness of the fading is also required
for assessing the quality of a link. This randomness can be assessed
by the Rician K-factor, as shown by physical fading channel modeling
\cite{Greenstein09}.

\paragraph*{Contributions of this work }

(\textit{i}) We design a MAC protocol where the AP assigns the partner
and transmission resources to each node based on information about
the path loss \emph{and} the Rician K-factor, extending the method
of \cite{Mahinthan08}. For this protocol we provide a performance
assessment in terms of network lifetime, utilizing realistic I2I/I2O
channel models.

(\textit{ii}) We propose an empirical, but analytically tractable,
stochastic model for the characterization of the two fading parameters,
namely, the path loss and the Rician K-factor for the links in an
indoor network with fixed nodes. The two parameters are modeled \textit{jointly}
generalizing existing scalar models, e.g., \cite{Greenstein09}. The
so-called \textit{bivariate }channel model is drawn and validated
using the multi-link channel measurement data \cite{Czink08}.

(\textit{iii}) For low-power operation of a cooperative WSN, we propose
a procedure where (1) the K-factor is estimated on a small number
of nodes with a slow update cycle: this distributed information is
conveyed to the central coordinator in order to update the bivariate
model parameters; (2) the estimation of the link-quality for each
transmission is based on the local average RSSI and the regularly
updated bivariate channel model as common a priori information. We
compare the performance of this approach with the one where both the
path-loss and the K-factor are permanently re-estimated for each transmission.\vspace{-3mm}

\section{System Model\label{system}}

The scenario under study consists of $N$ battery-powered indoor nodes
that communicate with a common AP located outdoors. The nodes transmit
during a communication session. In each session, the AP is acting
as the centralized coordinator for assigning the cooperating partners,
configuring the time-slot assignments and radio-frequency (RF) transmit
powers. Complex base-band notation is used to model the wireless link
between node $i\in\{1,\ldots,N\}$ and node $j\in\{0,\ldots,N\}$,
with node $j=0$ referring to the AP. The received signal at node
$j$ is $y_{i,j}=h_{i,j}x_{i}+n_{j}$ where $h_{i,j}$ is the frequency-flat
complex channel coefficient, $x_{i}$ is the symbol transmitted by
node $i$ with transmit power $\rho_{i}$, $n_{j}$ is additive symmetric
complex white Gaussian noise with variance $\sigma^{2}$. The instantaneous
signal-to-noise ratio (SNR) for from node $i$ to $j$ is modeled
as\begin{equation}
\gamma_{i,j}=\left(\rho_{i}/\sigma^{2}\right)\left\vert h_{i,j}\right\vert ^{2}.\label{eq:1-1}\end{equation}
The square fading envelope $\left\vert h_{i,j}\right\vert ^{2}$ is
constant for the whole codeword duration (block fading) and varies
from codeword to codeword according to the Rician distribution \cite{Erceg03},
such that $L_{i,j}=-\left(\mathbb{E}\left[\left\vert h_{i,j}\right\vert ^{2}\right]\right){}_{\mathrm{dB}}$
denotes the path loss in dB and $K_{i,j}=\left(\left|\mathbb{E}\left[h_{i,j}\right]\right|^{2}\right){}_{\mathrm{dB}}-\left(\mathrm{var}\left[h_{i,j}\right]\right){}_{\mathrm{dB}}$
denotes the Rician K-factor in dB. The term \textit{path loss}, used
for indicating $L_{i,j}$, includes the large-scale shadow fading
and the static component of the small-scale fading as detailed in
Sect. \ref{sub:Bivariate}. The block fading assumption motivates
the use of outage probability and it is confirmed by channel measurements
(see \cite{Czink08}). \vspace{-3mm}

\subsection{Medium Access Control and Node Coordination}

The transmission is organized into frames of duration $T_{\mathrm{F}}$,
further divided into $N+1$ subframes for time division multiple access
(TDMA), as shown in Fig. \ref{Flo: intro} (bottom). A unique subframe
of duration $T_{\mathrm{S}}=T_{\mathrm{F}}/(N+1)$ is assigned to
each node by the AP. The AP also provides the reference clock to all
the nodes, the grouping decisions and access coordination (e.g., power
and time slots allocation) through a periodic beacon transmission
\cite{Cal02}.

Let $\left(i,j\right)$ be a pair of cooperating nodes%
\footnote{We assume that each node can cooperate at most with one partner. The
extension of the analysis to grouping assignment with more than one
partner is beyond the scope of the paper.%
}, to accommodate cooperative transmission each subframe assigned to
any of these two nodes is further subdivided into two slots. As depicted
in Fig. \ref{Flo: intro}, for node $i$ the first slot spans a fraction%
\footnote{To simplify the mathematical treatment, we assume that the slot partition
$\beta$ can take any value in the interval $0<\beta<1$, although
in practice this is constrained to a finite number of data units. %
} $\beta_{i}=\beta$ of the subframe duration and it is used to transmit
the node $i$ data. The second slot with duration $\left(1-\beta_{i}\right)T_{\mathrm{S}}=\left(1-\beta\right)T_{\mathrm{S}}$
is reserved for helping the assigned partner node $j$. In a specular
way, the first and second slots of the subframe assigned to node $j$
span, respectively, the fractions $\beta_{j}=1-\beta$ (for node-$j$
data) and $\left(1-\beta_{j}\right)=\beta$ (for forwarding node-$i$
data). The AP optimally combines the noisy replicas of the signals
coming from the two nodes for data detection.\vspace{-3mm}

\subsection{Outage Probability Modeling\label{sub:Cooperation-gain}}

For \emph{point-to-point transmission} the outage probability is $P_{\mathrm{out}}^{\textrm{dir}}=\Pr[\gamma_{i,j}<\gamma_{\mathrm{th}}^{\mathrm{dir}}],$
where the SNR threshold $\gamma_{\mathrm{th}}^{\mathrm{dir}}=\left(2^{R}-1\right)/\Gamma$
, $R$ is the spectral efficiency measured in bit/channel use. The
gap $0\leq\Gamma\leq1$ can be varied by changing modulation/coding
format and targeted bit error rate (BER) level (see \cite{Wang03}
and references therein). The outage probability is assessed according
to the models in \cite{Savazzi09}, here adapted to the considered
scenario.

For the Rician fading model, the outage probability for node $i$
communicating directly with any node $j$ can be parametrized in terms
of the so-called coding gain $c_{i,j}$ \begin{equation}
P_{\mathrm{out}}^{\textrm{dir}}\approx\frac{\gamma_{\mathrm{th}}^{\mathrm{dir}}\sigma^{2}}{c_{i,j}\rho_{i}}\text{,}\label{app}\end{equation}
where $\approx$ indicates that the equality holds asymptotically
for high SNR%
\footnote{Tightness of the approximation is verified for SNR large enough to
guarantee sufficiently low outage probabilities ($\lesssim10^{-2}$).%
}. The coding gain $c_{i,j}$ depends on the K-factor $K_{i,j}$ and
the path loss $L_{i,j}$ according to: \begin{equation}
c_{i,j}=\frac{e^{\theta(K_{i,j})}}{\theta\left(L_{i,j}\right)\left[1+\theta\left(K_{i,j}\right)\right]}.\label{eq:coding_gain}\end{equation}
For convenience of notation we introduce the function $\theta(\cdot)=10^{(\cdot)/10}$
with inverse $\theta^{-1}(\cdot)=10\log_{10}(\cdot)$ (recall that
$K_{i,j}$ and $L{}_{i,j}$ have been defined in dB).

To model the performance of \emph{cooperative transmission} of node
$i$ with the help of node $j$ (here $j\neq0$), we consider the
AF relaying scheme\emph{ }as it has a simple architecture that facilitates
practical implementation. Node $j$ periodically overhears the signal
transmitted by the partner node $i$ and amplifies-and-forwards it
towards the AP \cite{Laneman04}. The amplification is based on a
variable gain approach, where the power amplification gain $a_{j}=\rho_{j}/\left[\sigma^{2}(\gamma_{i,j}+1)\right]$
is dynamically adjusted to the instantaneous SNR $\gamma_{i,j}$.
Note that the node must maintain a constant transmit power $\rho_{j}$
for the whole assigned subframe, i.e. also during the relaying phase,
so as to avoid amplifier non-linearities by switching the power level.
Recalling that the AP optimally combines the two noisy replicas of
the source signal, the effective SNR\begin{equation}
\gamma_{(i,j),0}=\gamma_{i,0}+\left(\frac{1}{\gamma_{i,j}}+\frac{1}{\gamma_{j,0}}+\frac{1}{\gamma_{i,j}\gamma_{j,0}}\right)^{-1}.\label{eq:snr}\end{equation}
 The outage probability for cooperative transmission \begin{equation}
P_{\mathrm{out}}^{\mathrm{coop}}=\Pr[\gamma_{(i,j),0}<\gamma_{\mathrm{th}}^{\mathrm{coop}}]\approx\frac{1}{2}\left(\frac{\gamma_{\mathrm{th}}^{\mathrm{coop}}\sigma^{2}}{c_{(i,j),0}\rho_{(i,j),0}}\right)^{d_{(i,j),0}},\label{eq:outage_gen}\end{equation}
with $\gamma_{\mathrm{th}}^{\mathrm{coop}}=\left(2^{R/\beta_{i}}-1\right)/\Gamma$.
The spectral efficiency $R$ is now multiplied by $1/\beta_{i}$ to
guarantee the same efficiency as for the non-cooperative case. Notice
that this assumption implies that the low-power radio transceiver
is designed to support multiple data rates \cite{TexasInstr10}. Also,
we select the same gap $\Gamma$ for all rates. Terms $\rho_{(i,j),0}$,
$c_{(i,j),0}$ and $d_{(i,j),0}$ denote the \emph{effective} power,
coding gain, and diversity order for the cooperative link $(i,j),0$,
respectively. For Rician fading, it is $d_{(i,j),0}=2$, $\rho_{(i,j),0}=\sqrt{\rho_{i}\rho_{j}}$
, and \begin{equation}
c_{(i,j),0}=\left[\frac{1}{c_{i,0}}\left(\frac{1}{c_{i,j}}+\frac{1}{c_{j,0}}\right)\right]^{-\frac{1}{2}},\label{eq:cAF}\end{equation}
where the coding gains depend on the K-factors and the path loss values,
as in (\ref{eq:coding_gain}).\vspace{-3mm}

\subsection{Energy Consumption Modeling\label{sub:Energy-Consumption-Modeling}}

The transmit power is designed so that the outage probability at the
AP is lower or equal to $p$. The corresponding energy consumption
for node $i$ during one frame is derived below.

\emph{No-cooperation} --\emph{ }From (\ref{app}), the transmit power
at node $i$ scales as\begin{equation}
\rho_{i}\approx\frac{\gamma_{\mathrm{th}}^{\mathrm{dir}}\sigma^{2}}{c_{i,0}}\frac{1}{p}.\label{eq:-4}\end{equation}
The average energy expenditure for node $i$ can thus be modeled as\begin{equation}
E_{i}=\rho_{i}T_{\text{S}}+E_{\text{RX}}\frac{T_{\text{S}}}{T_{\mathrm{F}}}+E_{\text{P}},\label{eq:-1}\end{equation}
where $E_{\text{RX}}$ is the energy consumption for receiving during
the beacon slot only, and $E_{\text{P}}$ is the energy consumption
for basic processing.

\emph{AF cooperation} -- For a small enough outage probability $p$,
it can be shown that the minimum transmit power levels for paired
nodes $(i,j)$ are\begin{equation}
\rho_{i}=\rho_{j}=\rho_{(i,j),0}=\kappa(\hat{\mathrm{\beta}})\sigma^{2}/\sqrt{2p},\label{eq:eq_power}\end{equation}
with $\kappa(\beta)=\max\left[\left(2^{R/\beta}-1\right)/\left(\Gamma c_{\left(i,j\right),0}\right),\:\left(2^{R/(1-\beta)}-1\right)/\left(\Gamma c_{\left(j,i\right),0}\right)\right]$.
The subframe fraction $\beta$ for the $i$-th node message is selected
so as to minimize $\rho_{(i,j),0}$: \begin{equation}
\hat{\beta}\approx\frac{1}{2}-\frac{\log_{2}(\lambda)}{8R},\label{eq:-7}\end{equation}
 with $\lambda=\sqrt{\frac{1+c_{i,0}/c_{i,j}}{1+c_{j,0}/c_{i,j}}}$
. The notation $\approx$ indicates that the scaling law is valid
for $\log_{2}(\lambda)\ll R$ (i.e., $\lambda\simeq1$). The proofs
of (\ref{eq:eq_power}) and (\ref{eq:-7}) are given in Appendix A.

Notice that using (\ref{eq:-7}) for $\lambda>1$ (i.e., $c_{i,0}>c_{j,0}$),
then $\hat{\beta}<1/2$ as the largest slot is reserved for helping
the partner $j$ that experiences more severe fading conditions. On
the other hand, if $c_{i,j}\gg\max\left[c_{i,0},\, c_{j,0}\right]$,
the slots have equal length, $\hat{\beta}\approx1/2$. Notice that
the choice $\beta=1/2$ is relevant also because it models a practical
system optimized for two data rates \cite{TexasInstr10}, i.e., the
lower for no-cooperation and the higher for cooperation.

The average consumed energy for node $i$ cooperating with partner
$j$ is then\begin{equation}
E_{(i,j),0}^{\mathrm{AF}}=\rho_{(i,j),0}T_{\text{\text{S}}}+E_{\text{RX}}(2-\beta)\frac{T_{\text{S}}}{T_{\mathrm{F}}}+(1+\upsilon_{\text{AF}})E_{\text{P}},\label{eq:en1}\end{equation}
 with $\upsilon_{\text{AF}}>0$ accounting for the energy consumption
for the partner signal amplification. The transmit power $\rho_{(i,j),0}$
for source and relay nodes are chosen as in (\ref{eq:eq_power}).\vspace{-3mm}

\section{Wireless Channel Characterization\label{modeling}}

In this section we present a framework for modeling and estimating
channel quality metrics for the links of the cooperative network.
The final aim is to provide metrics to be used at the MAC layer for
relay selection.

In Sect. \ref{sub:Bivariate} we propose a stochastic general model
of the two fading parameters $(L_{i,j},K_{i,j})$, which includes
as particular cases some models previously proposed in the literature
(see e.g. \cite{Greenstein09,Soma02}). In Sect. \ref{sub:Path-Loss K-factor},
the model is validated and discussed using experimental data collected
by an I2I/I2O measurement campaign, with indoor nodes deployed over
an office environment. The study will also provide a reliable simulation
environment for assessing the performance of partner selection algorithms
in Sect. \ref{results}. Given the a-priori knowledge of the model,
in Sect. \ref{sub:Maximum-Likelihood-(ML)-1} we propose a Bayesian
method for the estimation of the coding gain $c_{i,j}$ starting from
the path loss observation.\vspace{-3mm}

\subsection{Bivariate Model for the Large-Scale Fading Parameters\label{sub:Bivariate}}

We consider a fading channel $h_{i,j}$ between any two nodes $\left(i,j\right)$
of the static network. Link indexes $\left(i,j\right)$ will be omitted
when not needed, to simplify the notation. It is important to highlight
that, differently from mobile scenarios, here the deterministic channel
gain $\mu_{h}=E\left[h_{i,j}\right]$ accounts for the effects of
fixed scattering/absorbing objects which determine the so-called \emph{static
multipath} component. On the other hand, temporal fading - with variance
$\mathrm{\sigma_{h}^{2}=var}\left[h_{i,j}\right]$ - is due only to
some moving scatterers/absorbers in the environment. As assumed in
Sect. \ref{system}, this results in a Rician distribution with parameters
$L=-\theta^{-1}(\sigma_{h}^{2}+|\mu_{h}|^{2})$ and $K=\theta^{-1}(|\mu_{h}|^{2}/\sigma_{h}^{2})$
(recall that $\theta(\cdot)$ is the inverse of the transformation
to dB). The multipath configuration changes rapidly with the node
locations, thus leading to fast variations of the Rician factor $K$
and the path-loss $L$ over the space. The objective of this section
is the definition of a model to describe the variations of such parameters
from link to link.

Let $\mathbf{x}=[K\,\, L]^{\textrm{T}}$ be the vector collecting
the two fading parameters for a generic link. We model the variations
of $\mathbf{x}$ according to a bivariate Gaussian random variable
(bivariate model). We assume that $\mathbf{x}$ is Gaussian distributed,
$\mathbf{x}\sim\mathcal{N}\left(\mu_{\mathrm{\mathbf{x}}},\mathbf{C}\right)$,
with a mean $\mu_{\mathrm{\mathbf{x}}}(D)=[\ensuremath{\mu_{\mathrm{K}}(D)}\,\,\mu_{\mathrm{L}}(D)]^{\mathbf{\mathsf{\textrm{T}}}}$
depending on the link distance $D$ and a spatially invariant covariance
matrix \begin{equation}
\mathbf{C}=\mathbb{\left[\begin{array}{cc}
\mbox{\ensuremath{\textrm{\ensuremath{\sigma_{\mathrm{K}}^{2}}}}} & \varphi\sigma_{\mathrm{K}}\mbox{\ensuremath{\sigma_{L}}}\\
\varphi\mbox{\ensuremath{\sigma_{\mathrm{K}}}\mbox{\ensuremath{\sigma_{L}}}} & \mbox{\ensuremath{\sigma_{L}^{2}}}\end{array}\right]},\label{eq:bivariate-1}\end{equation}
The correlation coefficient $\varphi=\mathbb{E}\left[\left(K-\mu_{\mathrm{K}}(D)\right)\left(L-\mu_{\mathrm{L}}(D)\right)\right]/(\sigma_{\mathrm{K}}\mbox{\ensuremath{\sigma_{L}}})$
models the mutual dependence between the K-factor and the path loss
experienced over the same link. %
\begin{comment}
We assume that the temporal interval and the spatial area of observation
are chosen so as to guarantee stationarity up to the second order
channel statistics. %
\end{comment}
{}A metric, that will be relevant in partner selection analysis in Sect.
\ref{partnersel}, is the probability density function (pdf) of the
K-factor conditioned on the path loss, $\mathrm{p}\left(K|L\right)=\mathcal{N}(\mu_{\mathrm{K|L}},\mbox{ }\sigma_{\mathrm{K|L}}^{2})$,
with mean\begin{equation}
\mu_{\mathrm{K|L}}\mathrm{\mathbb{=E}}\left[K|L\right]=\mu_{\mathrm{K}}(D)+\frac{\textrm{\ensuremath{\sigma_{\mathrm{K}}}}}{\sigma_{\mathrm{L}}}\varphi\left(L-\mu_{\mathrm{L}}(D)\right)\label{eq:kfactor_cond}\end{equation}
and variance $\mbox{ }\sigma_{\mathrm{K|L}}^{2}=\mathrm{var}\left[K|L\right]=(1-\varphi^{2})\ensuremath{\sigma_{\mathrm{K}}^{2}}.$

The parameters of the bivariate model can be tuned for different propagation
scenarios, e.g. I2I and I2O, as done in the following.\vspace{-3mm}

\subsection{Experimental Calibration of the Bivariate Model and Analysis of Spatial
Coherence\label{sub:Path-Loss K-factor}}

In this section, we describe the calibration of the model parameters
$\mathbf{\mathbf{\mu}}_{\mathbf{x}}(D)$ and $\mathbf{C}$ on the
I2I and I2O propagation scenarios in Fig. \ref{Flo: intro}. We use
the multi-link channel measurements \cite{Czink08} at 2.45 GHz, but
also existing models whenever the data is not sufficient. It is important
to mention that the fading in the experiment is caused not only by
walking people but also by moving metallic objects. This contributes
to determine harsh I2O propagation conditions. The 70 MHz band of
the channel is divided into 60 subbands, each corresponding to a flat
fading subchannel. As mentioned in Sect. \ref{sub:Bivariate}, in
contrast to mobile scenarios, the small-scale temporal fading has
a different origin compared to the small-scale fading in the spatial
and spectral domains. Hence, $K_{i,j}$ and $L_{i,j}$ are estimated
independently in each subband.

\begin{figure}[t]
\centering{\includegraphics[width=15cm]{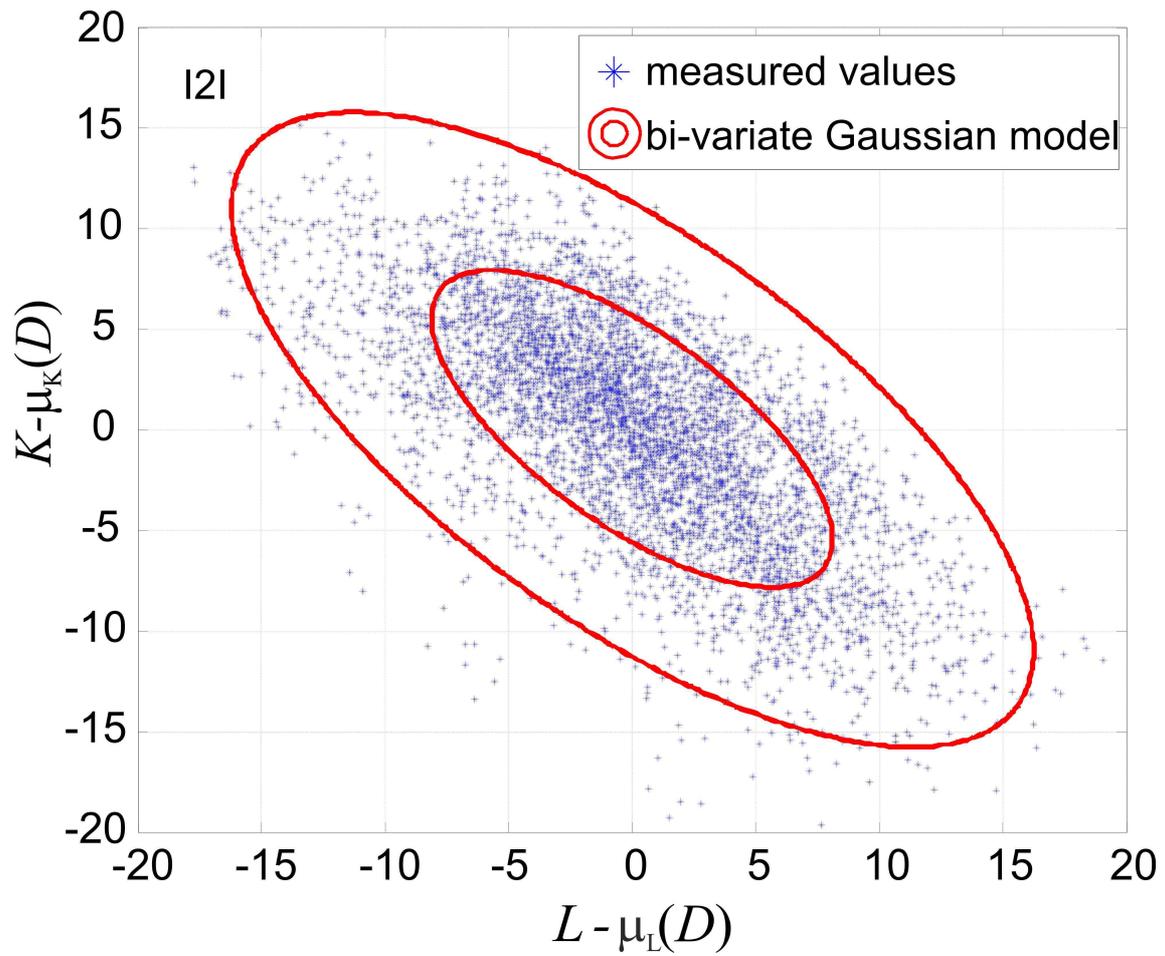}}

\caption{Measured I2I path loss and K-factor values minus the respective distance-dependent
means. The equidensity contours of the zero-mean bivariate Gaussian
distribution are depicted at one and two standard deviations from
the mean, containing 62\% and 98\% of the points respectively.}

\label{Flo:bivariate}%
\end{figure}

\emph{I2I channel model} -- The vector function $\mathbf{\mathbf{\mu}}_{\mathbf{x}}(D)$
is estimated by performing linear least squares regressions of $K$
and $L$ over the corresponding distances $D$ in logarithmic scale.
The covariance $\mathbf{C}$ is then obtained by computing variances
and covariances on the sets of data $\left\{ K-\mu_{\mathrm{K}}(D)\right\} $
and $\left\{ L-\mu_{\mathrm{L}}(D)\right\} $. The $l_{\infty}$ norm
of the error between the theoretic cumulative density function (cdf)
of $\left\{ K-\mu_{\mathrm{K}}(D)\right\} $ and $\left\{ L-\mu_{\mathrm{L}}(D)\right\} $
and the empirical one is 0.04, showing a very good fit. This is apparent
in Fig. \ref{Flo:bivariate}, where the equidensity contour lines
of the bivariate distribution of $(K-\mu_{\mathrm{K}}(D),L-\mu_{\mathrm{L}}(D))$
are shown together with the respective measured values.

\emph{I2O channel model} -- According to the urban micro-cell scenario
B4 in \cite{Kyosti07}, the propagation on the link $\left(i,0\right)$
is modeled as the combination of three main contributions: (\textit{i})
the indoor propagation from the node to the nearest wall to the BS
$\left(i\mathrm{,Wall}\right)$; (\textit{ii}) the propagation through
the wall; (\textit{iii}) the outdoor propagation from the wall to
the BS $\left(\mathrm{Wall}\mathrm{,0}\right)$. The overall link
path loss $L_{i,0}$ and K-factor $K_{i,0}$ are modeled as\begin{equation}
L_{i,0}=L_{i\mathrm{,Wall}}+L_{\mathrm{Wall}}+L_{\mathrm{Wall}\mathrm{,0}},\;\; K_{i,0}=K_{i\mathrm{,Wall}}+K_{\mathrm{Wall}\mathrm{,0}}.\label{eq:pathlossI2O}\end{equation}
Notice that the wall contribution has no effects on $K_{i,0}$. The
value chosen for the wall contribution is $L_{\mathrm{Wall}}=14\mathrm{dB}$
(neglecting the angle of the propagation path with respect to the
wall \cite{Kyosti07}). The specific value will not affect the performance
comparison of the algorithms presented in Sect. \ref{results}. The
outdoor parameters $\left(L_{\mathrm{Wall}\mathrm{,0}},K_{\mathrm{Wall}\mathrm{,0}}\right)$
modeling adheres to \cite[(4) and (9)]{Soma02}. The indoor parameter
$L_{i\mathrm{,Wall}}$ is modeled according to \cite[Table 4-4]{Kyosti07}.
The bivariate model $(L_{i\mathrm{,Wall}},K_{i\mathrm{,Wall}})$ and
the corresponding indoor correlation value are obtained by a simple
manipulation of the model for $L_{i\mathrm{,Wall}}$ in \cite[Table 4-4]{Kyosti07},
using the line slope and the error variance of the least squares linear
regression of $K$ over the corresponding $L$ in the available I2O
measurements. Numerical details are provided in Table 1.

The normal distribution of $K\mid L$ has been assumed by several
studies in the literature (see, e.g., \cite{Soma02,Erceg03,Greenstein09})
to model the residuals of the linear least square regression of $K$
over $L$. Here, instead, we have directly tested the bivariate normal
distribution of $\left\{ K-\mu_{\mathrm{K}}(D)\right\} $ and $\left\{ L-\mu_{\mathrm{L}}(D)\right\} $
via a 2-D fitting of the cdf. In this way, we have also highlighted
the strong (negative) correlation between $\left\{ K-\mu_{\mathrm{K}}(D)\right\} $
and $\left\{ L-\mu_{\mathrm{L}}(D)\right\} $, also observed experimentally
in our study (see the values detailed in Table 1). In contrast to
line-of-sight mobile scenarios, in our case the static multipath component
in $\mu_{h}$ mainly contributes to the negative correlation $\varphi$.
Consider e.g. two links $(i,j_{1})$ and $(i,j_{2})$, with nodes
$j_{1}$ and $j_{2}$ closely located, i.e., with distance in the
order of the carrier wavelength $\lambda$ ($D_{i,j_{1}}\approx D_{i,j_{2}}$).
Even if the two links are likely to experience the same channel shadowing,
average fading conditions $\mu_{\mathrm{\mathbf{x}}}(D)$, and degree
of temporal variations%
\footnote{This has been observed in the frequency domain also in \cite{Greenstein09}.
Notice that the small-scale spatial and frequency-selective fading
are caused by the same mechanism.%
} $\mathrm{\sigma_{h}^{2}}$, the two links still experience different
multipath configurations and thus different values for the static
multipath component in $\mu_{h}$. The variation on $\mu_{h}$ affects
with opposite sign the path-loss and the K-factor, thus, a \textit{strong}
negative correlation is observed.

\begin{figure}[t]
\centering{\includegraphics[width=15cm]{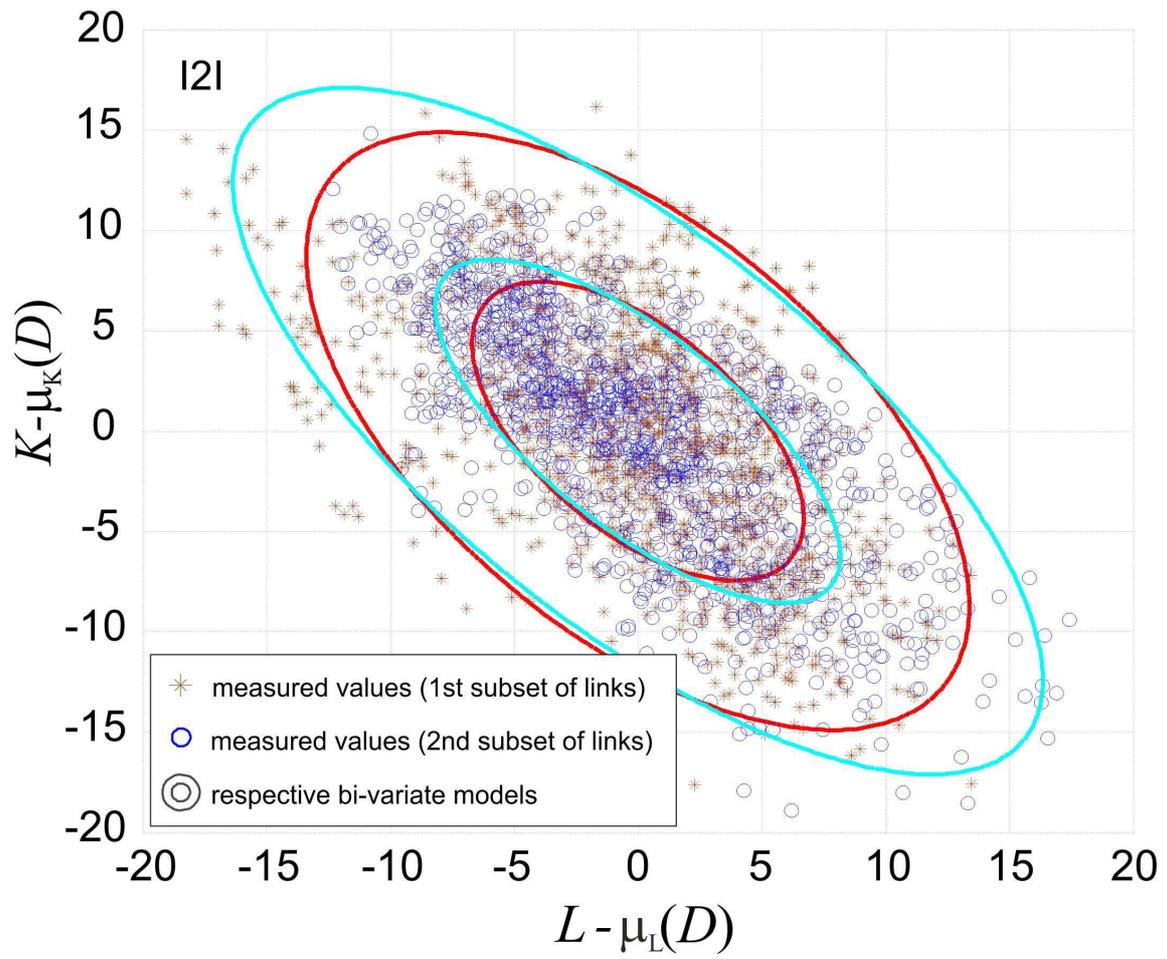}}

\caption{Measured I2I path loss and K-factor values minus the respective distance-dependent
means for the first and for the second subset of links, estimated
at two different time after 10 seconds, respectively. }

\label{Flo:bivariate_mismatch}%
\end{figure}

\emph{Spatial coherence of the models} -- We use the available measurement
data to empirically assess the spatial coherence of the I2I bivariate
model. We first estimate the model from the measurements over 16 links
during the \textit{first} $1/3$ of the total experiment duration
(32s). The model is again estimated from a disjoint set of 16 links
measured during the \textit{last} $1/3$ of the total experiment duration.
We observe that the model based on the first subset of links can predict
with extremely high accuracy the mean values $\mu_{\mathrm{\mathbf{x}}}(D)$
as modeled from the second subset of links. Fig. \ref{Flo:bivariate_mismatch}
shows the measured $\left(L-\textrm{\ensuremath{\mathcal{\mathcal{\mu_{\mathrm{L}}}}(D)}},K-\mathrm{\textrm{\ensuremath{\mathcal{\mu_{\mathrm{K}}}(D)}}}\right)$
values for the two subsets of links and the corresponding bivariate
Gaussian contour lines. The depicted bivariate models are strongly
matching as the covariance matrices $\mathbf{C}$ are practically
the same. Due to the common underlying physical mechanism, it is realistic
to assume that similar results are valid for the indoor propagation
model $(L_{i\mathrm{,Wall}},K_{i\mathrm{,Wall}})$ in the I2O case.
Finally, the outdoor $\left(L_{\mathrm{Wall}\mathrm{,0}},K_{\mathrm{Wall}\mathrm{,0}}\right)$
values, similarly to the wall penetration loss $L_{\mathrm{Wall}}$,
are not changing over the links and time in the considered I2O scenario
(so as to model a common stationary outdoor propagation scenario).
As analyzed in Sect. \ref{sub:Impact-of-imperfect}, these conclusions
suggest that the model parameters can be extracted efficiently employing
just a small subset of nodes.\vspace{-3mm}

\subsection{Link Quality Estimators\label{sub:Maximum-Likelihood-(ML)-1}}

In indoor environments, it is likely to incur into a link that exhibits
both a large average RSSI and, yet, a high packet loss rate \cite{Srinivasan06a}.
In these scenarios, accurate link quality estimation should include
a measure of the fluctuations of the received power. Based on the
physical fading channel modeling, several works propose ways to assess
the randomness of a link via the online estimation of its K-factor.
However, it is not trivial to predict the physical layer performance,
or even to draw MAC layer decisions directly using the estimated K-factor.

Let us instead consider the problem of the estimation of the coding
gain (\ref{eq:coding_gain}). It is convenient to re-write the coding
gain in dB scale as: \begin{equation}
c=\underbrace{\Upsilon\theta(K)-\theta^{-1}\left[1+\theta(K)\right]}_{\varsigma(K)}-L,\label{coding gain}\end{equation}
where \textbf{$\Upsilon=10\log_{10}(e)$}. The term $\varsigma(K)$
measures the additional information provided by the coding gain in
Rician fading $\left(\theta(K)>0\right)$ with respect to path-loss
information only.

Most commercial RF transceivers designed for low-power wireless applications
\cite{TexasInstr07,TexasInstr10} provide information about the received
signal strength from which the path-loss $L$ can be easily inferred
\cite{Srinivasan06a}. A straightforward method to estimate $c$ would
then be to calculate also the K-factor and to adopt the formula (\ref{eq:coding_gain}).
We refer to this estimate as \textit{direct} estimate $\hat{c}_{i,j}^{\mathrm{K}}$.
Accurate, but complex, estimators are proposed, e.g., in \cite{Tepedelenlioglu03}.
The most accurate estimator requires the knowledge of the fading phase.
On the other hand, for non-coherent transmissions as in current implementation
of IEEE 802.15.4 radios, the ratio between the squared mean and the
variance of the fading envelopes can be also used to estimate the
K-factor. Nevertheless, the estimation of small K-factor values ($K<3$dB)
becomes very inaccurate \cite{Koay06}.

Here, to reduce the estimation complexity, we propose to estimate
$c$ \textit{only} from the path loss $L$ by exploiting the a-priori
information on the statistics of $c|L$ derived from the bivariate
model. We assume the stationarity and the perfect knowledge of the
parameters $\mu_{\mathrm{\mathbf{x}}}(D)$ and $\mathbf{C}$ introduced
above. It is important to stress that the model knowledge is available
\textit{only} if the nodes are cooperating to exchange the information
required to extract the model parameters%
\footnote{The distributed estimation of the I2I and I2O models is beyond the
scope of the present work.%
}. The assumption of model knowledge is therefore realistic in the
considered network.

\begin{figure}[t]
\centering{\includegraphics[width=15cm]{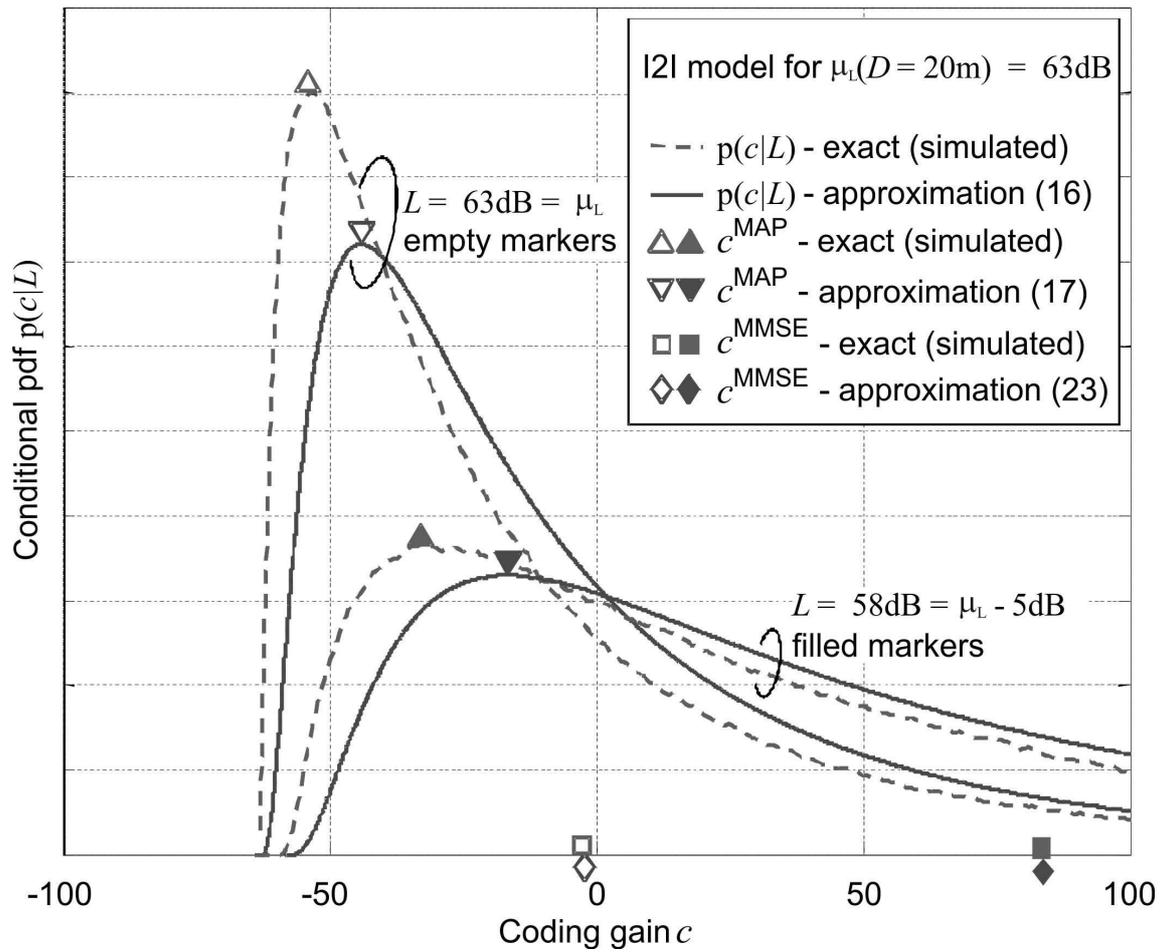}}

\caption{Conditional pdf $\mathrm{p}\left(c|L\right)$, both exact (simulated)
and approximated (analytical). The respective link quality estimators
$\hat{c}_{i,j}^{\mathrm{MAP}}$ and $\hat{c}_{i,j}^{\mathrm{MMSE}}$
are also depicted for two values of observed path loss $L$ (empty
markers for $L=63\mathrm{dB}$ and filled markers for $L=58\mathrm{dB}$).}

\label{Flo:pdfs}%
\end{figure}

Bayesian estimation of the link quality $c$ based on the observation
of $L$ requires the computation of the a-posteriori pdf $\mathrm{p}(c|L)$.
This pdf can be approximated by observing that for $\Upsilon\theta(K)\gg\theta^{-1}\left[1+\theta(K)\right]$
it is \textbf{$\varsigma(K)\approx\Upsilon\theta(K)$} and the link-quality
indicator (\ref{coding gain}) reduces to\textbf{ $c\approx\Upsilon\theta(K)-L$.
}Recalling that $K$ conditioned on the observed $L$ is Gaussian
distributed, $K\thicksim\mathcal{N}\left(\mu_{\mathrm{K|L}},\mbox{ }\sigma_{\mathrm{K|L}}^{2}\right)$,
it follows that $c$ is shifted log-normal distributed with pdf \cite{Walck00}:\begin{equation}
\begin{array}{l}
\mathrm{p}(c|L)\simeq\left[\left(\frac{c+L}{\Upsilon}\right)\sqrt{2\pi\sigma_{\mathrm{K|L}}^{2}}\right]^{-1}\exp\left(-\frac{1}{2\sigma_{\mathrm{K|L}}^{2}}\left[\theta^{-1}\left(\frac{c+L}{\Upsilon}\right)-\mu_{\mathrm{K|L}}\right]^{2}\right)u(c+L),\end{array}\label{eq:shift-logn}\end{equation}
where $u(x)$ is the unitary step function. The approximation is corroborated
by a numerical analysis in Fig. \ref{Flo:pdfs}, where the true and
the approximated pdfs are shown for two values of the path loss $L$
of one I2I link, simulated according to the model calibration in Sect.
\ref{sub:Path-Loss K-factor}.

Based on the a-posteriori pdf, we derive the \emph{maximum a posteriori}
(MAP) estimator of the link quality $c$. The MAP estimator, $\hat{c}^{\mathrm{MAP}}=\arg\underset{c}{\max}\left[\mathrm{p}(c|L)\right]$,
is approximated by the mode value of the shifted log-normal distribution
(\ref{eq:shift-logn}) which yields \cite{Walck00}:\begin{equation}
\begin{array}{l}
\hat{c}^{\mathrm{MAP}}\approx\Upsilon\theta\left(\mu_{\mathrm{K|L}}-\frac{\sigma_{\mathrm{K|L}}^{2}}{\Upsilon}\right)-L\end{array}.\label{map_est}\end{equation}

A \emph{minimum mean square error} (MMSE) estimator $\hat{c}^{\mathrm{MMSE}}$
is also provided in Appendix B. The MAP and MMSE estimators are depicted,
by empty markers for $L=63\mathrm{dB}$, and by filled ones for $L=58\mathrm{dB}$
in Fig. \ref{Flo:pdfs}, together with the corresponding pdfs. The
figure shows that the approximations of the pdfs and of the estimators
are tight to the exact ones, and behave similarly with varying path
loss value.\vspace{-3mm}

\section{Partner Selection Strategies\label{partnersel}}

In this section, the problem we tackle is how to select the partner
for the nodes that are tasked to communicate to the outdoor AP. The
final goal is to minimize the maximum energy consumed among them.
The AP selects the partners (relays) based on long-term link quality
metrics and not on the instantaneous channel gains, as proposed and
studied in the literature (e.g., in \cite{Ble06}). The optimal \textit{min-max}
pairing is found in Sect. \ref{sub:Optimal-pairing-for} that allows
also for configurations with un-paired nodes. To lower the complexity
and the amount of signaling, a worst-link-first (WLF) algorithm is
then described in Sect. \ref{sub:Worst-link-first-coding-gain-based-(WLF-CG)}.
Although the structure of the WLF algorithm is simple and resembles
the one in \cite{Mahinthan08} - with a modification for odd number
of nodes - the proposed performance analysis diverges substantially
for the I2I/I2O fixed links considered in our experimental scenario
(see Sect. \ref{modeling}). Given that the distributed wireless links
can be modeled by Rician fading with different K-factors, the first
key idea is that the AP uses the coding gain and not the path loss
as decision metric in the WLF algorithm. Secondly, the Bayesian estimators
derived in Sect. \ref{sub:Maximum-Likelihood-(ML)-1} provide a representation
of the link quality to be used by the AP to finalize the partner selection.\vspace{-3mm}

\subsection{Problem Definition and Optimal Solver\label{sub:Optimal-pairing-for}}

We define the set of candidate pairing sets $\mathcal{P}$, such that
one set $\xi\in\mathcal{P}$ contains up to $\left\lfloor N/2\right\rfloor $
\textit{disjoint} pairs of cooperative nodes: $\xi=\left\{ (i,j),(k,h),...,(f,g)\right\} .$
All the non-paired nodes belong to the set of single nodes $\mathcal{S_{\xi}}=\left\{ q,s,\ldots,z\right\} $,
such that $2\left|\mathcal{\xi}\right|+\left|\mathcal{S_{\xi}}\right|=N$
(where $\left|\cdot\right|$ denotes the cardinality of the set).
Given the candidate pairing set $\xi$ and the corresponding single
node set $\mathcal{S_{\xi}}$, the maximum energy consumed by a node
in the network is $E^{\mathrm{max}}(\xi)=\max[\max\limits _{(i,j)\in\xi}E_{(i,j),0}^{\mathrm{max}},\;\max\limits _{q\in\mathcal{S_{\xi}}}E_{q}]$,
where $E_{(i,j),0}^{\mathrm{max}}=\max[E_{(i,j),0},\, E_{(j,i),0}]$
is the maximum energy for the pair $(i,j)$ for a given relaying protocol,
with $E_{(i,j),0}$ defined by (\ref{eq:en1}). The optimal pairing
$\hat{\xi}$ is the solution to\begin{eqnarray}
\hat{\xi}=\arg\min_{\xi\in\mathcal{P}}E^{\mathrm{max}}(\xi) & ,\label{eq:optimal_sol}\end{eqnarray}
where we assume that all nodes have the same rate $R$ and outage
probability $p$ constraints.

The problem (\ref{eq:optimal_sol}) can be formulated as a special
case of the weighted matching problem on the non-bipartite graph $\mathcal{G=}(\mathcal{X},\mathcal{E})$
\cite{Papadimitriou98}. The set of vertices $\mathcal{X}$ corresponds
to the set of nodes $\left\{ 1,\ldots,N\right\} $, which are fully
connected by the set of undirected edges $\mathcal{E}=\left\{ e_{i,j}:\:\left(i,j\in\mathcal{X}\right)\:\&\:\left(i\leq j\right)\right\} $.
The loops $e_{i,j=i}$ can be regarded as edges $e_{i,\bar{i}}$,
where the virtual vertex $\bar{i}$ of the extended graph is connected
only to $i$. The weights $w\left(e_{i,j<i}\right)=E_{(i,j),0}^{\mathrm{max}}$
and $w\left(e_{i,j=i}\right)=E_{i}$ are associated to all the edges
and loops, respectively. The optimal pairing algorithm removes at
each iteration the maximum weighted edge of the extended graph (as
done in \cite{Chen07a} for a \textit{max-min} problem on the bipartite
graph) and checks the existence of a weighed matching solution in
the remaining graph using Gabow's algorithm%
\footnote{The Hungarian method, which is tailored for the bipartite graphs,
is instead considered in \cite{Chen07a}.%
} \cite[Ch. 11]{Papadimitriou98}, which was instead proposed in \cite{Mahinthan08}
for minimizing the sum of the energies consumed at the nodes in one
iteration only. It can be shown that the above algorithm reaches the
solution in $O\left(N^{5}\right)$ computational time. Notice that
the algorithm is centralized and requires the AP to know all the inter-node
link qualities for computing $c_{\left(i,j\right),0}$.\vspace{-3mm}

\subsection{Worst-Link-First Coding-Gain Based (WLF-CG) algorithm\label{sub:Worst-link-first-coding-gain-based-(WLF-CG)}}

The WLF algorithm is a suboptimal protocol for node pairing that allows
to reduce the complexity to $O\left(N^{2}\right)$. The conventional
WLF algorithm (referred to as WLF path-loss-based, WLF-PL) is based
on the information of second order statistics of the fading link \cite{Mahinthan08},
i.e. the path loss $L_{i,j}$. We propose a WLF method based on the
coding gain (WLF-CG).

\emph{Link quality estimation -- }Before pairing decisions can take
place, each node $i$ locally acquires an estimate of the link qualities
for all I2I links $\hat{c}_{i,j}$ to the candidate partners and the
I2O link towards the AP $\hat{c}_{i,0}$. Link quality estimator options
$\hat{c}_{i,j}^{\mathrm{K}}$, $\hat{c}_{i,j}^{\mathrm{MAP}}$, and
$\hat{c}_{i,j}^{\mathrm{MMSE}}$ (in dB) have been discussed in Sect.
\ref{sub:Maximum-Likelihood-(ML)-1}. The WLF-PL algorithm uses $\hat{c}_{i,j}=-L_{i,j}$.

\emph{Protocol structure -- }The algorithm is composed of two phases:

\textit{1) Candidate partner set discovery --} Each node $i$ performs
link quality estimation $\hat{c}_{i,0}$ for the I2O channel exploiting
the periodic transmission of beacon subframes from the AP as probing
signals to be used for channel parameter estimation. The link qualities
$\hat{c}_{i,j}$ measured from all neighboring nodes over I2I links
are estimated based on the signals overheard from the potential partners.
The difference $\hat{c}_{i,j}-\hat{c}_{i,0}$ is compared at each
node $i$ to a common threshold $\tau$ in order to guarantee the
condition $\theta(\hat{c}_{i,j})\gg\theta(\hat{c}_{i,0})$. If $\hat{c}_{i,j}-\hat{c}_{i,0}>\tau$
then node $j$ becomes a candidate partner for node $i$. Given the
I2O link quality estimation $\hat{c}_{i,0}$, the threshold $\tau$
is centrally designed such that the probability of finding no candidate
partners among the $N-1$ potential candidates $\prod_{j=1,j\neq i}^{N}\left[1-\Pr\left(c_{i,j}>\hat{c}_{i,0}+\tau|L_{i,j}\right)\right]$
is small enough, where $\Pr\left(c_{i,j}>\hat{c}_{i,0}+\tau|L_{i,j}\right)$
can be estimated using (\ref{eq:shift-logn}). The evaluation of the
optimal value for the threshold $\tau$ is carried out in the case-study
outlined in Sect. \ref{results}. The candidate partners set $\mathcal{C_{P}}(i)=\{j\,:\,\hat{c}_{i,j}-\hat{c}_{i,0}>\tau\}$
is finally communicated to the AP from each node $i$, using the assigned
subframe.

\textit{2) Assignment algorithm at the AP --} At each iteration the
AP selects the worst-uplink node $i$ with link quality $\hat{c}_{i,0}<\hat{c}_{k,0}\mbox{ }$,$\forall k\neq i$,
and, if possible, assigns it to the best-uplink candidate partner
$j$, such that\begin{equation}
j=\arg\max_{j\in\mathcal{C_{P}}(i)}\ensuremath{\mbox{ }\hat{c}_{j,0}}\label{eq:candidate}\end{equation}
Nodes $i$ and $j$ are paired and disregarded in the next iterations,
unless $\mathcal{C_{P}}(i)$ is empty. In the latter case, node $i$
is left un-paired in the final configuration.

For an odd number of nodes $N$ the preliminary step is to leave the
best-uplink node un-paired, then the partner assignment follows as
above for the remaining $N-1$ nodes. \vspace{-3mm}

\section{Experimental Assessment of Partner Selection in the I2O Environment\label{results}}

In what follows, we provide numerical simulations on the performance
of the partner selection algorithms (Sect. \ref{partnersel}) and
of the Bayesian link quality estimation methods (Sect. \ref{sub:Maximum-Likelihood-(ML)-1}).
The network setup is outlined in Fig. \ref{Flo: intro}, $N$ nodes
are randomly distributed in a $25\mathrm{m}\times25\mathrm{m}$ indoor
environment, while the AP is placed outdoors $50\mathrm{m}$ away
from the nearest wall. For each random topology of the network, K-factor
and path loss values are generated independently for all links according
to the respective I2I and I2O models detailed in Sect. \ref{sub:Path-Loss K-factor}.
The performance results are expressed in terms of \textit{lifetime
gain} $\mathbb{E}[E^{\mathrm{max}}(\xi_{\mathrm{ref}})]/\mathbb{E}[E^{\mathrm{max}}(\hat{\xi})]$,
defined as the ratio between the maximum energies (averaged over $5\times10^{4}$
random topologies) consumed among the nodes according to two different
pairing strategies: the reference protocol with pairing solution $\xi_{\mathrm{ref}}$
and the proposed protocol with pairing solution $\hat{\xi}$. Note
that the largest values of $E^{\mathrm{max}}(\xi_{\mathrm{ref}})$
and of $E^{\mathrm{max}}(\hat{\xi})$ highlight the lifetime gain
in scenarios where the I2O channel conditions are worst. In the examples,
the energies consumed for reception $E_{\text{RX}}$ and for basic
processing $E_{\text{p}}$ are neglected. According to the I2O modeling
in Sect. \ref{modeling}, it is likely that $c_{i,j}\gg\max\left[c_{i,0},\, c_{j,0}\right]$.
From (\ref{eq:-7}), the optimal choice for slot duration is $\beta_{i}=\beta_{j}=1/2$.
We set $\Gamma=1$, as this assumption has no relevant impact on the
performance comparisons in the examples. The target outage probability
is $p=10^{-3}$ for all the nodes with spectral efficiency $R=1\,\mathrm{bps/Hz}$.

In Sect. \ref{sub:WLF-CG-Protocol-Performance}, we provide numerical
simulations for the WLF-CG lifetime performance. In Sect. \ref{sub:Impact-of-link},
the impact of the proposed link quality estimation is evaluated and
compared to the case where a noisy estimation of the K-factor is obtained
from training. Finally, in Sect. \ref{sub:Impact-of-imperfect}, the
proposed pairing algorithm performance is discussed in presence of
imperfect modeling of the channel. \vspace{-3mm}

\subsection{WLF-CG Protocol Performance\label{sub:WLF-CG-Protocol-Performance}}

\begin{figure}
\centering{\includegraphics[width=15cm]{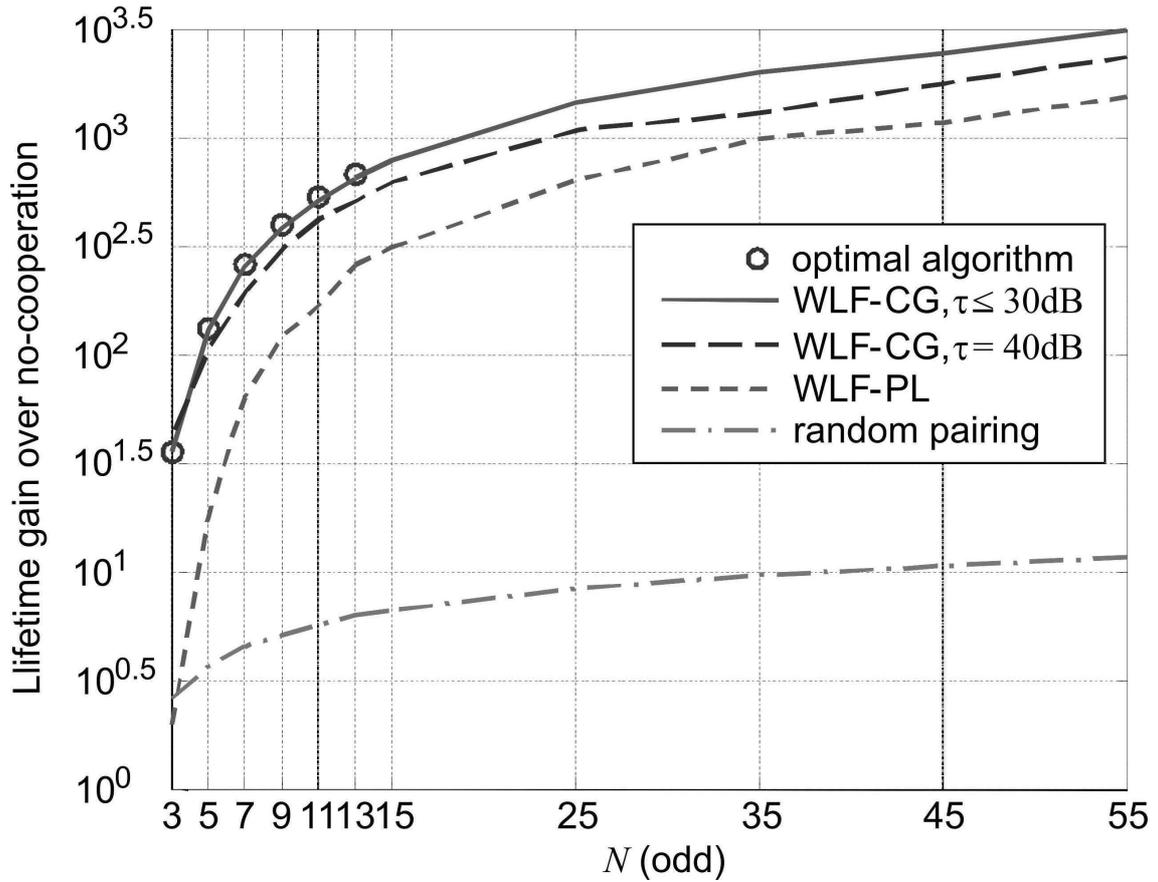}}

\caption{Cooperative transmission lifetime gain over no-cooperation with different
partner selection strategies and varying number of transmitting nodes
$N$ (odd values). The WLF-CG uses threshold values $\tau=\{30\mathrm{dB},40\mathrm{dB}\}$
for $\hat{c}_{i,j}-\hat{c}_{i,0}$ in the candidate partner set discovery
phase. The conservative choice $\tau=-\infty$ is used for the WLF-PL.}
\label{Flo:odd_gains}%
\end{figure}

We first evaluate the performance of the WLF-CG algorithm assuming
that the link qualities $c_{i,j}$ are perfectly known at the respective
nodes $i=\{1,\ldots,N\}$ and $j=\{0,\ldots,N\}$. Fig. \ref{Flo:odd_gains}
shows the lifetime gain of AF cooperation compared to no-cooperation,
i.e. $\xi_{\mathrm{ref}}=\oslash$ is the empty pairing set and $\hat{\xi}$
is the pairing set obtained according to a partner selection strategy.
A random pairing strategy is also considered for comparison where
all nodes are disjointly paired with a random choice of the partner.
For odd $N$, the optimal pairing algorithm in Sect. \ref{sub:Optimal-pairing-for},
the random pairing strategy, the WLF-PL algorithm, and the proposed
WLF-CG are compared. The candidate partner conditions (see Sect. \ref{sub:Worst-link-first-coding-gain-based-(WLF-CG)})
$c_{i,j}\gg c_{i,0}$ and $L_{i,j}\ll L_{i,0}$ for the WLF-CG and
the WLF-PL, respectively, are almost always guaranteed for each pair
of nodes, in particular it is $\Pr\left(c_{i,j}>c_{i,0}+\tau\right)\simeq1$
and $\Pr\left(L_{i,j}<L_{i,0}-\tau\right)\simeq1$ for $\tau\leq30\mathrm{dB}$.
For the propagation environment under consideration, the simulations
show that partner selection performance get worse when choosing $\tau\geq40\mathrm{dB}$.
The exploitation of the knowledge of the K-factor is revealed crucial:
the WLF-CG algorithm increases the lifetime from a factor of 20 for
$N=3$ to a factor of 2 for $N=55$ compared to the WLF-PL. This results
from the fact that the WLF-CG algorithm allows for a more efficient
exploitation of the available diversity, as if the optimal algorithm
in Sect. \ref{sub:Optimal-pairing-for} were applied. The remarkable
gains over no-cooperation and over the random pairing denote a large
degree of spatial redundancy provided by the multi-link channel, i.e.,
the path loss and K-factor values exhibit significant variations over
the space.\vspace{-3mm}

\subsection{Impact of Link Quality Estimation on WLF-CG\label{sub:Impact-of-link}}

The optimality of the WLF-CG pairing strategy relies on the accuracy
of coding gain (link quality) estimation. This motivates a closer
study to identify the most suitable estimator and to quantify the
benefits provided by the knowledge of the distributed channel model.
Here, we assume that only the path loss $L$ is perfectly known.

\begin{figure}
\centering{\includegraphics[width=15cm]{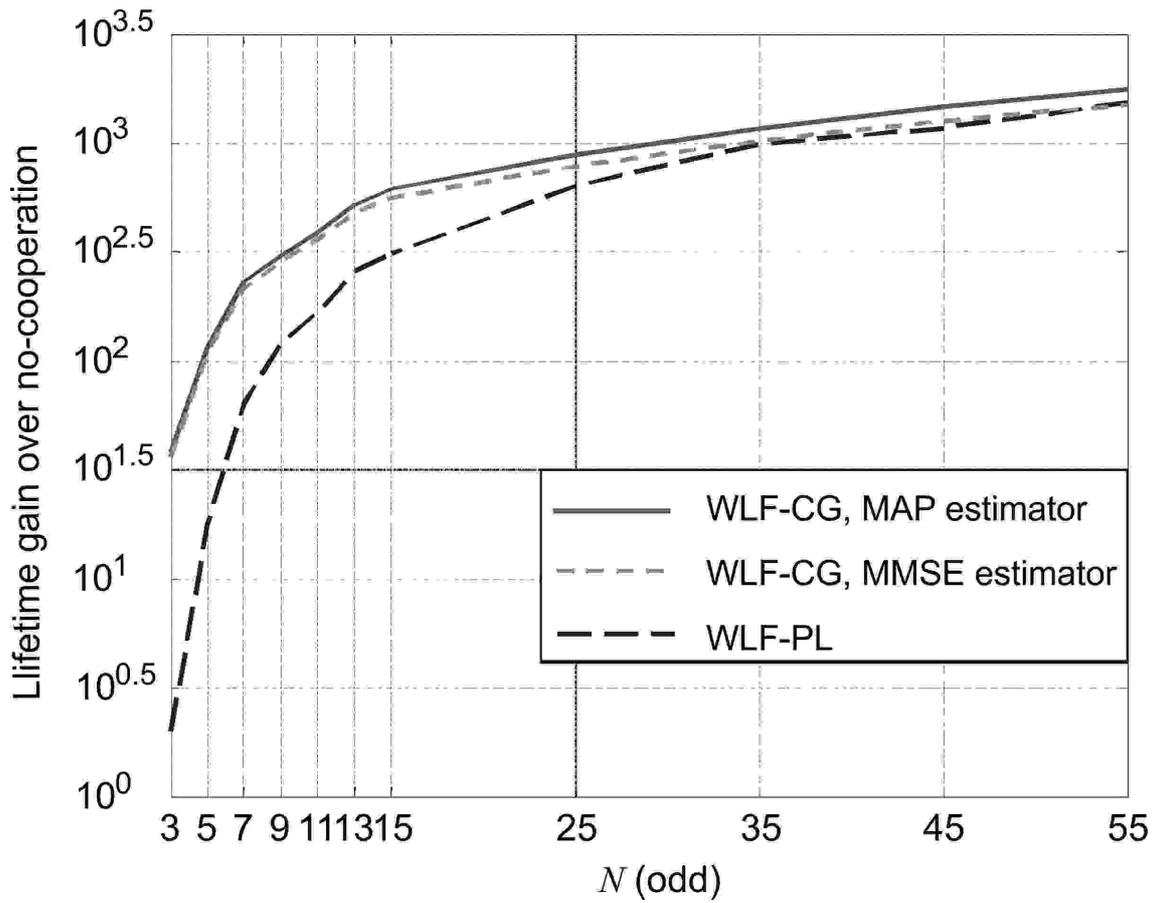}}

\caption{The lifetime gains of the WLF-CG algorithm based on $\hat{c}_{i,j}^{\mathrm{MAP}}$
and $\hat{c}_{i,j}^{\mathrm{MMSE}}$ over no-cooperation, compared
to that of WLF-PL, with varying number of transmitting nodes $N$
(odd values).}
\label{Flo:odd_gains-est}%
\end{figure}

\emph{Link quality estimation from the path loss }$\hat{c}_{i,j}=\left\{ \hat{c}_{i,j}^{\mathrm{MAP}},\hat{c}_{i,j}^{\mathrm{MMSE}}\right\} $
-- The WLF-CG algorithm can capitalize from the available site-specific
channel characterization. Fig. \ref{Flo:odd_gains-est} shows the
energy gains of cooperation over no-cooperation for varying number
of cooperating nodes $N$. The estimators $\hat{c}_{i,j}^{\mathrm{MAP}}$
and $\hat{c}_{i,j}^{\mathrm{MMSE}}$ are used, exhibiting equivalent
performance: notably, the lifetime gain over the WLF-PL ranges between
factors 2 ($N=15$) and 20 ($N=3$). These gains are similar to those
obtained in the case where $c_{i,j}$ are perfectly known (see Fig.
\ref{Flo:odd_gains}). Thus, the proposed Bayesian estimation of $c$
is revealed useful to guarantee significant lifetime benefits in the
considered network settings.

\begin{figure}
\centering{\includegraphics[width=15cm]{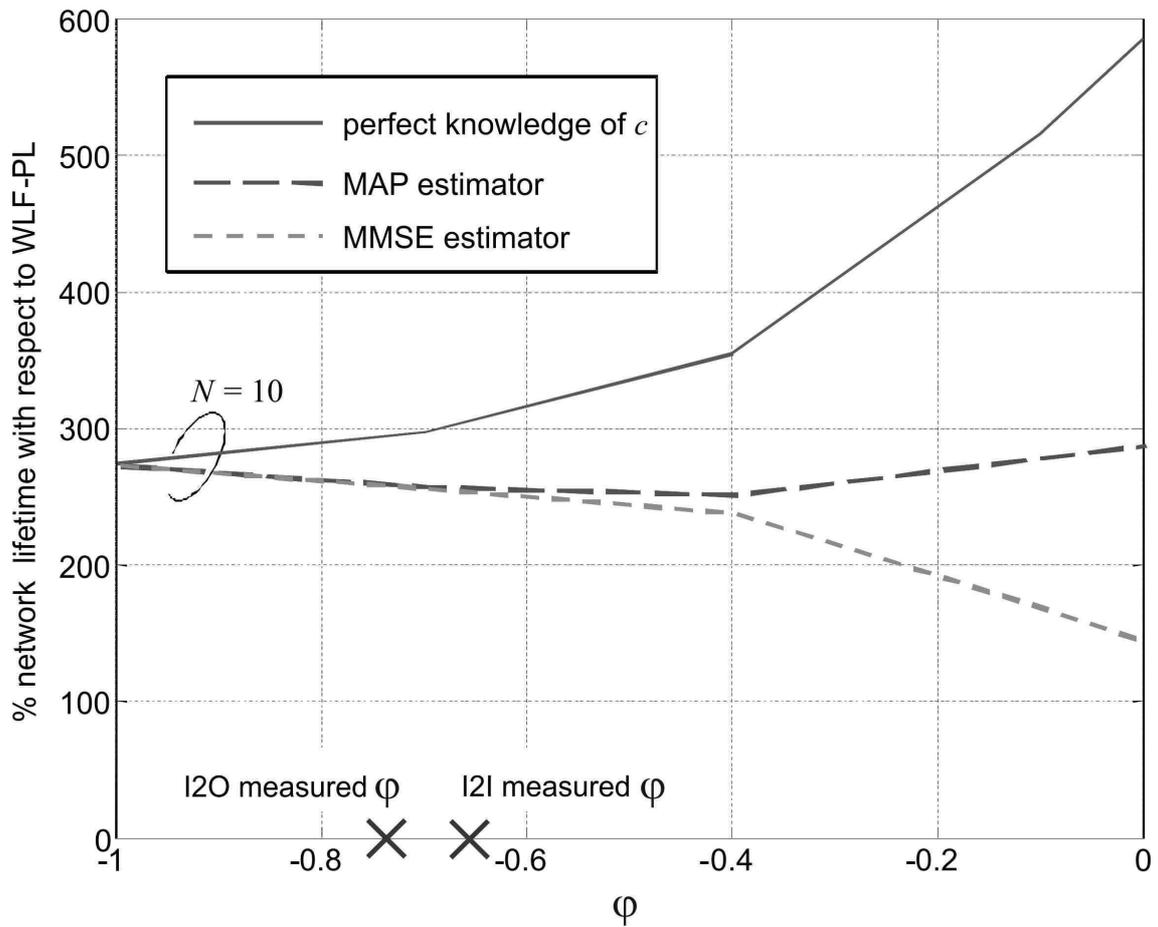}}

\caption{Lifetime performance of the WLF-CG algorithm, also with MAP and MMSE
estimators, compared in percentage to the one of the WLF-PL (100\%
means equal performance). The correlation $\varphi$ between path
loss and K-factor as defined in (\ref{eq:bivariate-1}) is varying
and is assumed equal for both the I2I links and the indoor component
of the I2O links. As practical reference, the cross markers highlight
the values $\varphi=-0.66$ and $\varphi=-0.74$, i.e., the correlation
observed in the I2I and in the indoor component of the I2O measurements,
respectively.}

\label{Flo:corr_vary}%
\end{figure}

In Fig. \ref{Flo:corr_vary} we evaluate how the WLF-CG lifetime gain
scales with the correlation $\varphi$ between the path loss and the
K-factor for $N=10$. The correlation $\varphi$ (assumed equal for
both the I2I links and the indoor component of the I2O links) varies
arbitrarily from $-1$ to $0$, whereas the other channel parameters
conform to the values in Table 1. When the link quality $c$ is known,
the lifetime gain increases as $\varphi$ gets closer to 0: this is
the case where the knowledge of the K-factor becomes in theory most
beneficial. Instead, the WLF-CG algorithm based on $\hat{c}_{i,j}^{\mathrm{MAP}}$,
and $\hat{c}_{i,j}^{\mathrm{MMSE}}$ is shown not to be sensitive
within realistic variations $-0.8<\varphi<-0.5$, as it improves always
by a factor 2.5 the lifetime obtained by employing the WLF-PL. It
is interesting to observe that $\hat{c}_{i,j}^{\mathrm{MAP}}$ performs
better than $\hat{c}_{i,j}^{\mathrm{MMSE}}$ for $\varphi>-0.4$,
as it is more conservative in predicting the worst-uplink link quality,
which dominates the lifetime performance. Indeed, it can be verified
that $\hat{c}_{i,j}^{\mathrm{MAP}}\leq\hat{c}_{i,j}^{\mathrm{MMSE}}$.

\begin{figure}
\centering{\includegraphics[width=15cm]{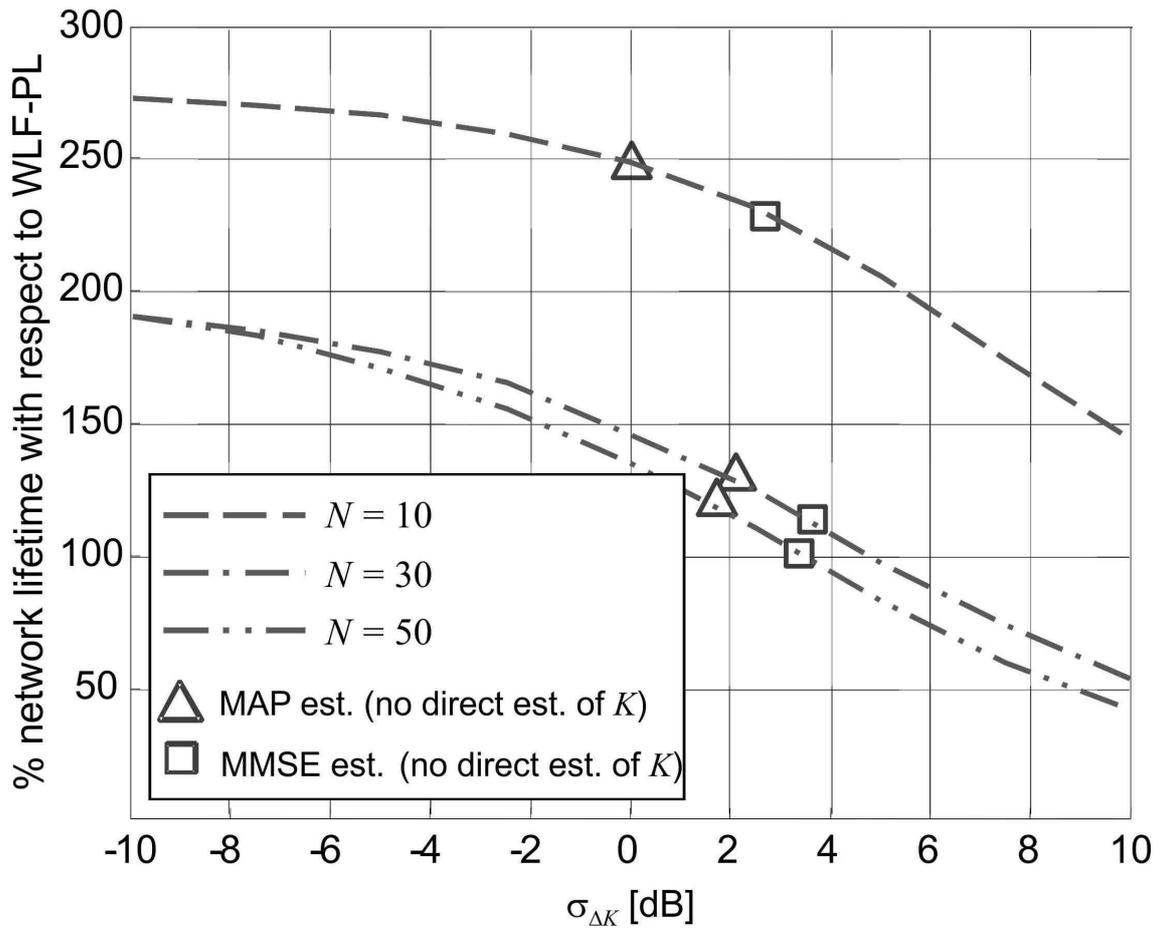}}

\caption{WLF-CG lifetime compared in percentage to WLF-PL, with varying K-factor
estimation MSE. The lifetime performance of the WLF-CG algorithm based
on $\hat{c}_{i,j}^{\mathrm{MAP}}$ and $\hat{c}_{i,j}^{\mathrm{MMSE}}$
are also marked on the respective curves for comparison.}

\label{Flo:Kerr}%
\end{figure}

\emph{Link quality estimation from both path loss and K-factor }$\hat{c}_{i,j}=\hat{c}_{i,j}^{\mathrm{K}}$
--\emph{ }The estimator of the K-factor can be obtained from the complex
fading realizations or the corresponding squared envelope values (acquired
from RSSI measurements) \cite{Tepedelenlioglu03}. As discussed in
Sect. \ref{sub:Maximum-Likelihood-(ML)-1}, depending on the choice
of the estimator, various trade-offs between accuracy and complexity
can be obtained. Here, we prefer not to consider any specific estimator
$\hat{K}$, but we rather model the estimation noise $\theta\left(\hat{K}\right)-\theta\left(K\right)$
as zero-mean Gaussian, where the pdf is truncated in order to keep
$\theta\left(\hat{K}\right)$ positive. Fig. \ref{Flo:Kerr} shows
the lifetime gain of the WLF-CG algorithm with the direct estimate
$\hat{c}_{i,j}=\hat{c}_{i,j}^{\mathrm{K}}$ over the WLF-PL, with
varying root mean squared error (RMSE) $\sigma_{\bigtriangleup K}=\sqrt{\mathbb{E}\left[\left|\theta\left(\hat{K}\right)-\theta\left(K\right)\right|^{2}\right]}$.
Remarkably, for $N=10$ and $\sigma_{\bigtriangleup K}\leq5\mathrm{\mathrm{d}B}$
the lifetime is at least doubled. Notice that $\hat{c}_{i,j}^{\mathrm{MAP}}$
outperforms $\hat{c}_{i,j}^{\mathrm{MMSE}}$, resulting in lifetime
gains as if $\hat{c}_{i,j}^{\mathrm{K}}$ were employed with $\sigma_{\bigtriangleup K}=0\mathrm{\mathrm{d}B}$,
and $3\mathrm{\mathrm{d}B}$, respectively. The estimator of $\hat{c}_{i,j}^{\mathrm{K}}$
is revealed robust for partner selection for a small number of cooperating
nodes $N$. Instead, for larger number of nodes $N\geq30$, the lifetime
is doubled only for $\sigma_{\bigtriangleup K}\leq-10\mathrm{\mathrm{d}B}$,
while the WLF-CG is even outperformed by the WLF-PL for $\sigma_{\bigtriangleup K}>6\mathrm{dB}$.
\vspace{-3mm}

\subsection{\emph{Impact of Imperfect Model Knowledge\label{sub:Impact-of-imperfect}}}

We consider a protocol where a set of $N_{1}$ nodes is used in a
prior communication session to update the channel model, as shown
in Fig. \ref{Flo: intro}. In a later communication session, $N_{2}$
different nodes are then tasked to transmit to the AP. The impairment
of the model knowledge is due to the limited number of the $(K,L)$
regression points used in the prior session to estimate the model,
i.e., $(N_{1}^{2}-N_{1})/2$ points for the I2I model and $N_{1}$
points for the I2O model. Thus, the accuracy of the proposed Bayesian
link quality estimation improves for increasing $N_{1}$, at the expense
of spectral and computational power efficiency. This trade-off is
assessed in the following. We focus on the MAP estimator $\hat{c}_{i,j}^{\mathrm{MAP}}$,
that was shown to have better performance compared to $\hat{c}_{i,j}^{\mathrm{MMSE}}$
in the above evaluations.

\begin{figure}
\centering{\includegraphics[width=15cm]{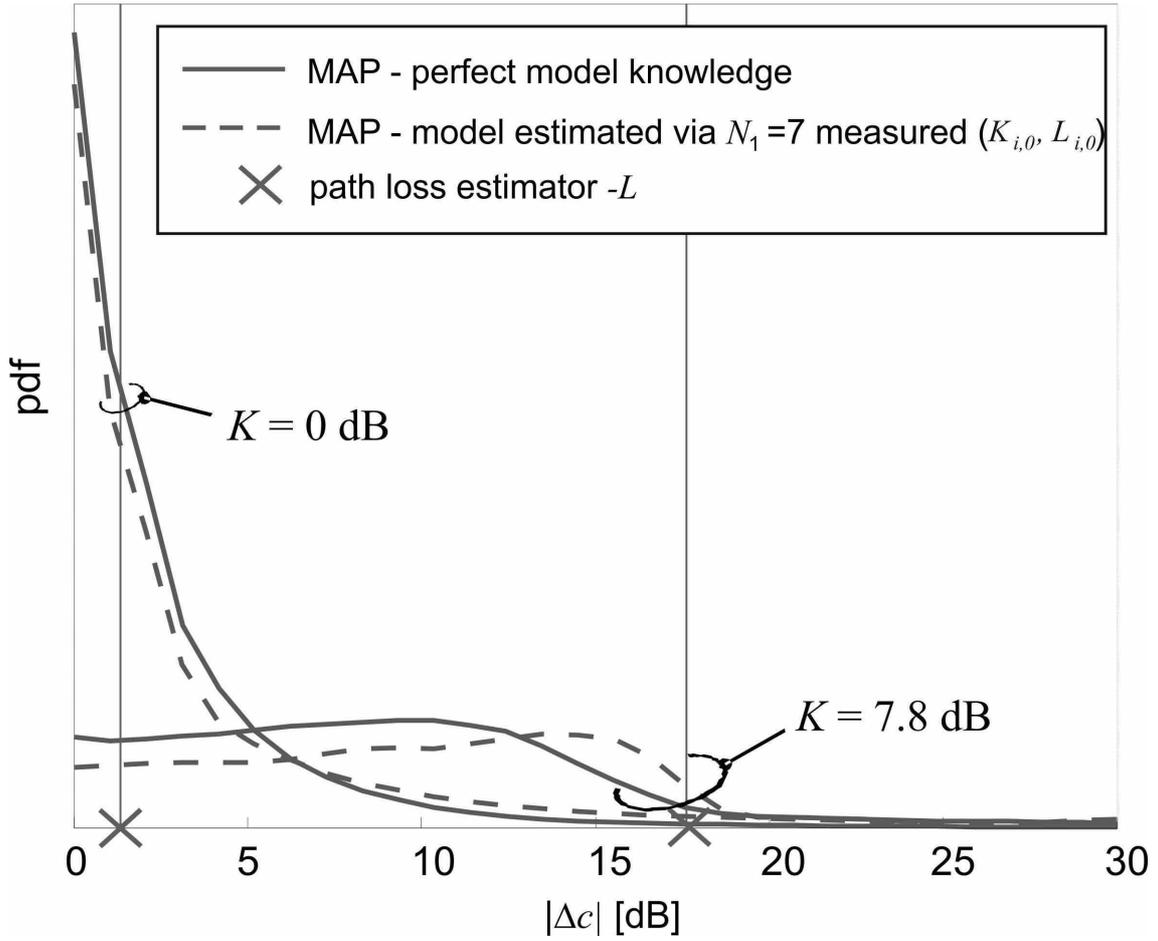}}

\caption{I2O coding gain absolute estimation error $\left|\Delta c\right|=\left|\hat{c}_{i,0}-c_{i,0}\right|$
pdf for the estimator $\hat{c}_{i,0}=\hat{c}_{i,0}^{\mathrm{MAP}}$
with perfect model knowledge (solid lines) compared to that with the
model estimated via $N_{1}=7$ points ($L_{i,0}$,$K_{i,0}$) observed
in the previous communication session (dashed lines). Also the estimation
error $\varsigma(K)=c_{i,0}+L_{i,0}$ in the WLF-PL is depicted for
the considered values of the K-factor $K_{i,0}=\{0\mathrm{dB},7.8\mathrm{dB}\}$,
i.e., $\theta(K_{i,0})=\{1,6\}$.}

\label{Flo: C_err_pdf}%
\end{figure}

Fig. \ref{Flo: C_err_pdf} shows the pdfs of the I2O coding gain estimation
absolute error in dB $\Delta_{c}=|\hat{c}_{i,0}-c_{i,0}|$ for the
estimator $\hat{c}_{i,0}^{\mathrm{MAP}}$ with perfect (solid lines)
and imperfect model knowledge derived via $N_{1}=7$ nodes (dashed
lines) for $K_{i,0}=\{0\mathrm{dB},7.8\mathrm{dB}\}$. Notably, the
absolute error pdfs are very similar, although the bivariate model
is estimated by only 7 spatially distinct measurements of $(K_{i,0},L_{i,0})$.
As expected, the mean points of $\Delta_{c}$ in the considered range
of K-factor values, i.e., >0dB and <7.8dB, are smaller than the absolute
error of the link quality estimator used in the WLF-PL, i.e., $\varsigma(K)=c_{i,0}+L_{i,0}$
(depicted by the cross markers).

\begin{figure}
\centering{\includegraphics[width=15cm]{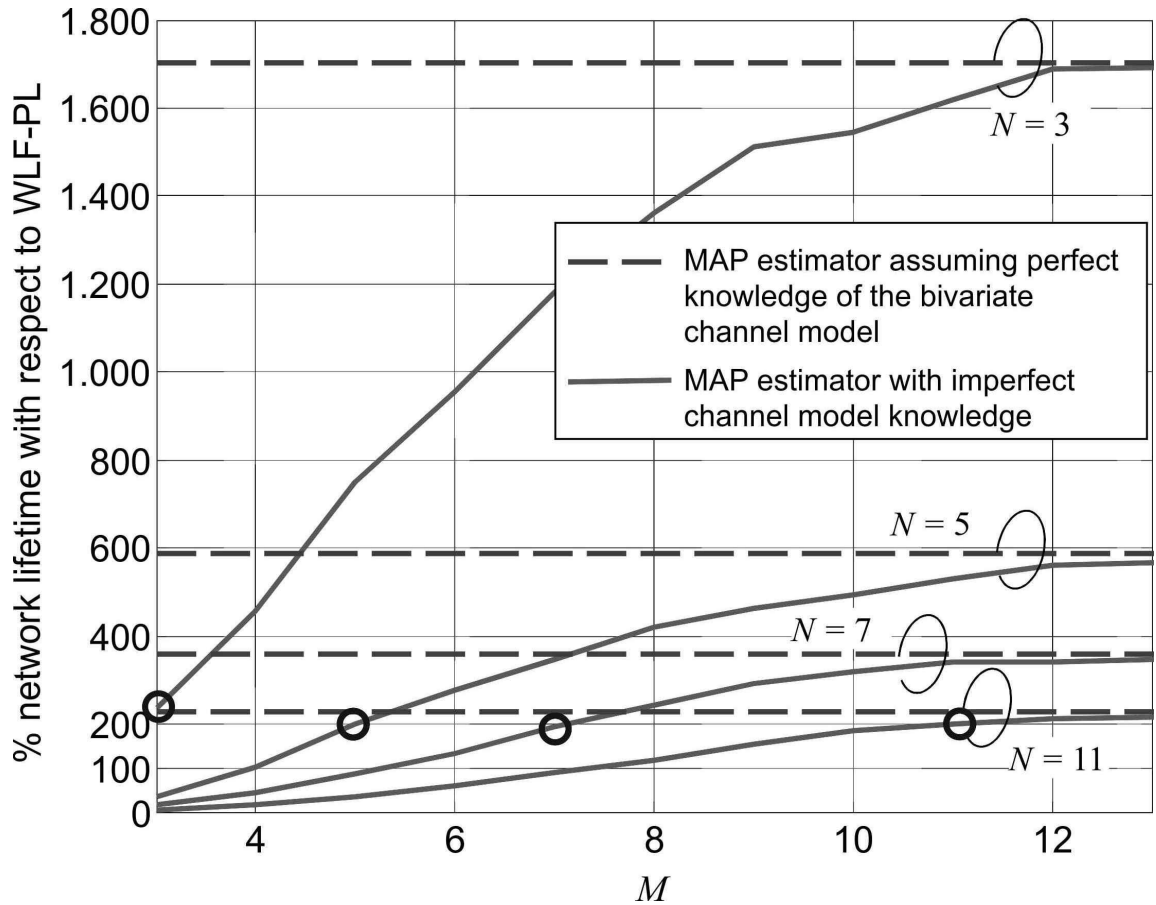}}

\caption{WLF-CG lifetime with MAP link-quality estimator $\hat{c}_{i,j}^{\mathrm{MAP}}$
compared in percentage to WLF-PL, with varying number of nodes $N_{1}$
involved in the the estimation the bivariate channel model during
the previous communication session. The lifetime gains of the WLF-CG
algorithm based on $\hat{c}_{i,j}^{\mathrm{MAP}}$ with perfect model
knowledge are also depicted with dashed lines as upper bound. Performance
for different values of communicating nodes $N_{2}$ in the current
session are depicted. The circle markers highlight the performance
for $N_{1}=N_{2}$.}

\label{Flo: Anchors-2}%
\end{figure}

Fig. \ref{Flo: Anchors-2} plots the network lifetime gains of the
WLF-CG that uses $\hat{c}_{i,j}=\hat{c}_{i,0}^{\mathrm{MAP}}$ over
the WLF-PL with varying number of the nodes $N_{1}$ employed for
the parameter extraction. The performance with perfect model knowledge
is also depicted as upper bound. It is revealed that, for $3<N_{2}<11$,
the proposed WLF-CG algorithm outperforms the WLF-PL, by \textit{at
least} doubling the network lifetime for $N_{1}\geq N_{2}$ (as highlighted
by circle markers). \vspace{-5mm}

\section{Concluding Remarks}

We have designed an efficient partner selection protocol for cooperative
wireless access from fixed indoor nodes to an outdoor AP through AF
relaying. Given that the considered links can be modeled by Rician
fading, we have proposed a partner selection algorithm that adopts
the coding gain - a function of the path loss and K-factor - as link
quality metric for pairing the nodes. The proposed approach implies
an additional computational cost due to the estimation of the K-factor,
but provides a network lifetime increase by factors ranging from 2
to 20 compared to the conventional algorithm based on path loss information
only.

Analyzing measurement data from a channel sounding campaign at 2.4GHz
we were able to characterize the path loss and the K-factor with a
Gaussian bivariate model. From this bivariate model a novel link quality
indicator is derived that does not require a permanent re-estimation
of the K-factors. This is the Bayesian estimate of the coding gain,
where the path loss is the observed variable (inferred through RSSI
readings) and the channel model is the a-priori information. The novel
metric improves remarkably the partner selection performance almost
as if full knowledge of the K-factors were available.

Furthermore, the analysis on the measurement data reveals the high
degree of spatial coherence of the model. Hence, we have proposed
a protocol with two phases: (\textit{i}) for a long time period (within
the stationarity time interval of the model) low-complexity communication
sessions take place, where the link qualities are inferred through
path loss measurements and the a-priori information of the channel
model; (\textit{ii}) for a short time period, a more complex communication
session occurs, where a set of nodes estimate also the K-factors in
order to update the site-specific channel model. Numerical results
show a good trade-off between performance and robustness. In particular,
the proposed protocol allows to double the network lifetime compared
to the conventional algorithm, also in presence of modeling mismatches.\vspace{-3mm}

\section*{Appendix A}

For Rician fading where $\rho_{(i,j),0}=\sqrt{\rho_{i}\rho_{j}}$
and $d_{(i,j),0}=2$, the target outage probability $p$ constrains
the powers $\rho_{i}$ and $\rho_{j}$ over the two slots such that
repetition based coding prescribes that $\beta_{i}=\beta$ and $1-\beta_{j}=\beta$:\vspace{-2mm}
\textbf{\begin{equation}
\begin{array}{c}
\rho_{j}\geq\frac{1}{2}\left[\frac{\sigma^{2}\left(2^{R/(1-\beta)}-1\right)}{\Gamma c_{\left(j,i\right),0}}\right]^{2}\left(p\rho_{i}\right)^{-1},\;\rho_{i}\geq\frac{1}{2}\left[\frac{\sigma^{2}\left(2^{R/\beta}-1\right)}{\Gamma c_{\left(i,j\right),0}}\right]^{2}\left(p\rho_{j}\right)^{-1}.\end{array}\label{eq:pwdConstraints}\end{equation}
}Recall that the gap $\Gamma$ can be designed to be the same for
the communication both from $i$ and from $j$. By minimizing the
maximum over $\rho_{i}$ and $\rho_{j}$, the simple power balancing
solution $\rho_{i}=\rho_{j}=\kappa(\hat{\mathrm{\beta}})\sigma^{2}/\sqrt{2p}$
is found\textbf{ }where $\kappa(\beta)=\max\left[\left(2^{R/\beta}-1\right)/\left(\Gamma c_{\left(i,j\right),0}\right),\:\left(2^{R/(1-\beta)}-1\right)/\left(\Gamma c_{\left(j,i\right),0}\right)\right]$
and $\hat{\mathrm{\beta}}=\arg\underset{\beta}{\min}\kappa(\beta)$
is solution to $\left(2^{R/\hat{\beta}}-1\right)c_{\left(j,i\right),0}=\left(2^{R/\left(1-\hat{\beta}\right)}-1\right)c_{\left(i,j\right),0},$
therefore\begin{equation}
2^{R/\hat{\beta}}-1=\lambda\left(2^{R/\left(1-\hat{\beta}\right)}-1\right)\label{eq:-6}\end{equation}
where $\lambda=\sqrt{\frac{1+c_{i,0}/c_{i,j}}{1+c_{j,0}/c_{i,j}}}$
. The solution to (\ref{eq:-6}) is now approximated for large enough
rate $R$ such that for $\log_{2}(\lambda)\ll R$ it is \begin{equation}
\frac{R}{\hat{\beta}}-\frac{R}{1-\hat{\beta}}\approx\log_{2}\lambda.\label{eq:soll}\end{equation}
Now by letting $\hat{\beta}=\frac{1}{2}-\hat{\upsilon}$ with $\hat{\upsilon}$
small enough, the solution to (\ref{eq:soll}) is $\hat{\upsilon}\approx\frac{\log_{2}(\lambda)}{8R}$.
\vspace{-3mm}

\section*{Appendix B}

The \emph{minimum mean square error} (MMSE) estimator $\hat{c}^{\mathrm{MMSE}}$
can be approximated as\begin{eqnarray}
\hat{c}^{\mathrm{MMSE}} & = & \mathbb{E}\left[c|L\right]\approx\mathbb{E}\left[\Upsilon\theta(K)|L\right]-\mathbb{E}\left[\max\left(K,0\right)|L\right]-L=\nonumber \\
 & = & \Upsilon\theta\left(\mu_{\mathrm{K|L}}+\frac{\sigma_{\mathrm{K|L}}^{2}}{2\Upsilon}\right)-\mu_{\mathrm{K|L}}Q\left(-\frac{\mu_{\mathrm{K|L}}}{\sigma_{\mathrm{K|L}}}\right)-\frac{\sigma_{\mathrm{K|L}}}{\sqrt{2\pi}}\exp\left(-\frac{\mu_{\mathrm{K|L}}^{2}}{2\sigma_{\mathrm{K|L}}^{2}}\right)-L,\label{mmse_est}\end{eqnarray}
where $Q\left(\cdot\right)$ is the Q-function. In (\ref{mmse_est})
we use the approximation \textbf{$\varsigma(K)\simeq\Upsilon\theta(K)-\max\left(K,0\right)$}
$\forall K$%
\footnote{For $K\gg0$dB it is $\theta^{-1}\left[1+\theta(K)\right]\approx K$,
for $K\ll0$dB $\theta^{-1}\left[1+\theta(K)\right]\approx0$dB. Values
around 0dB are not significant for the mean evaluation.%
}, where $\mathbb{E}\left[\max\left(K,0\right)|L\right]=\intop_{0}^{\infty}K\,\mbox{p}\left[K|L\right]\,\mbox{d}K$.\vspace{-3mm}

\bibliographystyle{ieeetr}
\bibliography{IEEEabrv,Zemen2}

\pagebreak{}\newpage{}\clearpage{}\thispagestyle{empty}%
\pagebreak{}\newpage{}\clearpage{}\thispagestyle{empty}%
\pagebreak{}\newpage{}\clearpage{}\thispagestyle{empty}%
\begin{table}[b]
\centering{}\caption{Parameters for the Bivariate Model (\ref{eq:bivariate-1}) - see Sect.
\ref{sub:Path-Loss K-factor}}
\begin{tabular}{|c|c|}
\cline{2-2}
\multicolumn{1}{c|}{} & Bivariate Model (\ref{eq:bivariate-1}) Parameters\tabularnewline
\hline
$\begin{array}{c}
\mbox{Indoor-to-Indoor (I2I)}:\\
(i,j)\neq0;\: D=D_{i,j}\end{array}$ & $\begin{array}{c}
\\\mu_{\mathbf{x}}\begin{cases}
\mathcal{\mu_{\mathrm{K}}}(D|_{\mathrm{m}})=16.90-10\alpha_{K}\log_{10}\left(D|_{\mathrm{m}}\right)\\
\mathcal{\mu_{\mathrm{L}}}(D|_{\mathrm{m}})=40.4+10\alpha_{\mathrm{L}}\log_{10}\left(D|_{\mathrm{m}}\right)\end{cases}\\
\\\left[\alpha_{\mathrm{K}}=0.53,\:\alpha_{\mathrm{L}}=1.75\right]\\
\\\mathbf{C}\begin{cases}
\varphi=-0.66\\
\mbox{ }\sigma_{\mathrm{K}}=5.8\mathrm{dB},\mbox{ }\sigma_{\mathrm{L}}=6\mathrm{dB}\end{cases}\\
\\\end{array}$\tabularnewline
\hline
$\begin{array}{c}
\mathrm{\mbox{Indoor-to-Outdoor (I2O),}}\\
\mbox{Outdoor propagation \ensuremath{K_{\mathrm{Wall}\mathrm{,0}}\mbox{ and }L_{\mathrm{Wall}\mathrm{,0}}}:}\\
(\mathrm{wall},\: j=0\mbox{ });\: D=D_{\mathrm{wall},0}\end{array}$ & $\begin{array}{c}
\\\mu_{\mathbf{x}}\begin{cases}
\mathcal{\mu_{\mathrm{K}}}(D|_{\mathrm{km}})=7.85-10\alpha_{K}\log_{10}\left(D|_{\mathrm{km}}\right)\\
\mathcal{\mu_{\mathrm{L}}}(D|_{\mathrm{km}})=135.78+10\alpha_{\mathrm{L}}\log_{10}\left(D|_{\mathrm{km}}\right)\end{cases}\\
\\\left[\alpha_{\mathrm{K}}=0.45,\:\alpha_{\mathrm{L}}=3.89\right]\\
\\\mathbf{C}\begin{cases}
\varphi=-0.25\\
\mbox{ }\sigma_{\mathrm{K}}=7.5\mathrm{dB},\mbox{ }\sigma_{\mathrm{L}}=7.9\mathrm{dB}\end{cases}\\
\\\end{array}$\tabularnewline
\hline
$\begin{array}{c}
\mathrm{\mbox{Indoor-to-Outdoor (I2O),}}\\
\mbox{Indoor propagation \ensuremath{K_{i,\mathrm{Wall}}\mbox{ and }L_{i,\mathrm{Wall}}}:}\\
(i\neq0,\mbox{ }\mathrm{wall});\: D=D_{i,\mathrm{wall}}\end{array}$ & $\begin{array}{c}
\\\mu_{\mathbf{x}}\begin{cases}
\mathcal{\mu_{\mathrm{K}}}(D|_{\mathrm{m}})=-0.3D|_{\mathrm{m}}\\
\mathcal{\mu_{\mathrm{L}}}(D|_{\mathrm{m}})=0.5D|_{\mathrm{m}}\end{cases}\\
\\\mathbf{C}\begin{cases}
\varphi=-0.74\\
\sigma_{\mathrm{K}}=5.7\mathrm{dB},\mbox{ }\sigma_{\mathrm{L}}=7\mathrm{dB}\end{cases}\\
\\\mbox{ }\end{array}$\tabularnewline
\hline
\end{tabular}%
\end{table}
\pagebreak{}\newpage{}\clearpage{}\thispagestyle{empty}%
\pagebreak{}\newpage{}\clearpage{}\thispagestyle{empty}%
\pagebreak{}\newpage{}\clearpage{}\thispagestyle{empty}

\pagebreak{}\newpage{}\clearpage{}\thispagestyle{empty}%

\pagebreak{}\newpage{}\clearpage{}\thispagestyle{empty}%
\pagebreak{}\newpage{}\clearpage{}\thispagestyle{empty}%
\pagebreak{}\newpage{}\clearpage{}\thispagestyle{empty}%
\pagebreak{}\newpage{}\clearpage{}\thispagestyle{empty}%

\end{document}